 \documentclass[smallabstract,smallcaptions]{dccpaper}

\usepackage[english]{babel}
\usepackage{graphicx}
\usepackage{amsthm}
\usepackage{multirow}
\usepackage[ruled,vlined]{algorithm2e}
\usepackage{caption}
\usepackage{subcaption}
\usepackage{url}
\usepackage{cite}
\usepackage{amssymb}
\usepackage{bbm}
\usepackage{algorithmicx}
\usepackage{algpseudocode}
\usepackage{mathtools}
\usepackage{amsfonts}
\usepackage{cases}
\usepackage{float}
\usepackage{booktabs}
\usepackage{xcolor}
\usepackage{tikz}
\usepackage[commandnameprefix=always]{changes}
\setcommentmarkup{\textcolor{magenta}{\,[#1]}}
\usepackage{verbatim}
\usepackage{tabularx}
\usepackage{capt-of}
\usepackage[hidelinks]{hyperref}
\usepackage[capitalize,nameinlink]{cleveref}
\usepackage{comment}

\newcommand{\proofversion}{2}

\ifnum\proofversion=2
  \includecomment{vlong}
  \excludecomment{vshort}
\fi

\graphicspath{{figures/}}
\newcommand{\bp}{\mathbf{p}}

\newcommand{\pu}{\mathbf{p}_{u}}
\newcommand{\bq}{\mathbf{q}}
\newcommand{\bu}{\mathbf{u}}
\newcommand{\bv}{\mathbf{v}}
\newcommand{\bw}{\mathbf{w}}
\newcommand{\bx}{\mathbf{x}}
\newcommand{\by}{\mathbf{y}}
\newcommand{\bz}{\mathbf{z}}
\newcommand{\be}{\mathbf{e}}
\newcommand{\bn}{\mathbf{n}}
\newcommand{\bone}{\mathbf{1}}
\newcommand{\ceil}[1]{\left\lceil #1\right\rceil}

\newcommand{\Sint}{\mathring{\mathbb{S}}}

\newtheorem{theorem}{Theorem}
\newtheorem{corollary}[theorem]{Corollary}
\newtheorem{lemma}[theorem]{Lemma}
\newtheorem{definition}[theorem]{Definition}
\newtheorem{remark}[theorem]{Remark}
\newtheorem{proposition}[theorem]{Proposition}

\newlength{\figurewidth}
\newlength{\smallfigurewidth}

\begin{document}

\title
{\large
\textbf{Quantization and Codebook Stability of Euclidean and KL Lattice Quantizers}
}

\author{%
Haoxuan Luo$^{\ast}$, Ryan Gabrys$^{\dag}$, and Farzad Farnoud$^{\ast}$%
\thanks{This work was supported in part by the NSF under grants CIF-2312871 and CAREER-2144974. This work has been accepted for poster presentation at the 2027 Data Compression Conference (DCC).}\\[0.5em]
{\small\begin{minipage}{\linewidth}\begin{center}
\begin{tabular}{ccc}
$^{\ast}$Electrical and Computer Engineering & \hspace*{0.5in} & $^{\dag}$Calit2 \\
University of Virginia && University of California--San Diego \\
Charlottesville, VA, USA && San Diego, CA, USA\\
\url{{ayr7tb,farzad}@virginia.edu} && \url{ryan.gabrys@gmail.com}
\end{tabular}
\end{center}\end{minipage}}
}

\maketitle
\thispagestyle{empty}

\begin{abstract}

\begin{vlong}
Small edits to raw data can cause large changes in the compressed version, which can be costly in applications such as data synchronization and incremental backup. We study quantization stability, the probability that two perturbed distributions map to the same representative, and codebook stability, the probability that the resulting Huffman codebooks are the same. We first quantize the empirical distribution and then construct a Huffman code from the quantized distribution, which can improve codebook stability at the cost of some compression efficiency.

We study two quantizers based on a shifted and scaled lattice in the probability simplex: the Euclidean Lattice Quantizer (ELQ) and the KL Quantizer (KLQ). For ELQ, we derive bounds on quantization stability and KL distortion. For KLQ, we show that its quantization cells can be determined by comparisons with adjacent lattice codewords and use the distances to the cell boundaries to derive a quantization stability bound. For a symmetric codebook of positive types, we obtain the exact worst-case KL distortion over the full probability simplex. We also give a heuristic choice $\alpha^\star$ of the minimum codeword probability and compare it with the minimum coordinate of a Lloyd-optimized codebook. Simulations using canonical Huffman coding compare ELQ and KLQ with an existing compander-based quantizer and show the trade-off between KL distortion, codebook stability, and encoded length under random deletions.
\end{vlong}

\end{abstract}

\section{Introduction}
\label{sec:intro}

\begin{vlong}
Lossless compression reduces storage and communication costs while allowing the original data to be recovered exactly. Methods such as Huffman coding and arithmetic coding \cite{Huffman,arithmetic} use source probabilities to reduce the number of bits needed to encode the data. In incremental backups and data synchronization, a small change to a file may still require a large update to its compressed version \cite{synchronization,rsync}. This can increase the amount of data that needs to be transferred or rewritten. For Huffman coding, a small edit can change the empirical distribution and cause many codewords to change when the code is rebuilt \cite{ITWpaper}. Even symbols that were not edited may then be assigned different codewords, increasing the number of bits that change in the compressed version.

Smooth compression asks that small edits to the input cause only small changes in the compressed version. Montanari and Mossel \cite{montanari2008smooth} defined smoothness using a Lipschitz condition on Hamming distance and constructed smooth codes based on sparse graphs. Vatedka and Tchamkerten \cite{vatedka2020local} studied a related problem of local decoding and update, where a small part of the source can be read or updated by accessing only a small number of bits in the compressed data. Our previous work \cite{ITWpaper} focused on Huffman coding and studied how its codebook changes under edits. We showed that canonical Huffman coding is more stable than standard Huffman coding and proposed a pre-allocation method that further improves stability while keeping the Huffman code optimal. Varshney et al. \cite{varshney2009malleable} studied another setting that trades some compression efficiency for a smaller edit distance between compressed versions.

One way to make this trade-off is to quantize the source distribution before coding. Our recent work \cite{ISITpaper} studied this approach. Nearby empirical distributions may map to the same quantized distribution and use the same codebook. We proposed a weighted geometric quantizer and a covering-code method over the type space, and studied their KL distortion and code stability under edits. Probability distribution quantization has also been studied independently of smooth compression. Reznik \cite{reznik2011} studied Euclidean nearest-neighbor quantization on type lattices and biased types whose entries remain positive. Geiger and B\"ocherer \cite{geiger2016greedy} studied $M$-type approximation and gave an exact greedy algorithm for minimizing KL divergence over all $M$-types. Adler et al. \cite{adler2023efficient} studied a scalar quantization method for probability distributions and analyzed its KL distortion.

\begin{table}[!b]
\centering
\caption{Comparison of KL-distortion results with prior work.}
\label{tab:kl_comparison}
\renewcommand{\arraystretch}{1.2}
\begin{tabular}{@{}p{0.27\linewidth}p{0.67\linewidth}@{}}
\toprule
Setting & KL distortion \\
\midrule

\shortstack[l]{\textbf{ELQ}\\[-1mm]
unclipped cell,\\[-1mm]
$\min_i q_i\ge\eta>0$}
&
\begin{tabular}[t]{@{}l@{}}
Prior: $D=O(n^{-2})$ \cite{reznik2011} \\[1mm]
Ours: $\displaystyle D\le\frac{m\delta^2}{4\eta}$
\end{tabular}
\\[3mm]

\shortstack[l]{\textbf{KLQ}\\[-1mm]
$p\in\Delta^\circ_{m-1},$\\[-1mm]
$p\to e_j$}
&
\begin{tabular}[t]{@{}l@{}}
Prior upper bound:
$\displaystyle
\to\ln\frac{K}{K-m+1}
+\frac{\ln 2}{2(K-m+1)}
$
\cite{geiger2016greedy}
\\[1mm]
Ours:
$\displaystyle
D\to\ln\frac{K}{K-m+1}
$
\end{tabular}
\\

\bottomrule
\end{tabular}
\end{table}

In this paper, we study the stability and KL distortion of lattice quantizers. Since the same quantized distribution gives the same Huffman codebook, we focus on quantization stability in our analysis. We consider two quantizers on a shifted and scaled $A_{m-1}$ lattice inside the probability simplex, whose codebook is a special case of the biased types studied by Reznik~\cite{reznik2011}. The Euclidean Lattice Quantizer (ELQ) uses Euclidean distance, which can be bounded by TV distance. The KL Quantizer (KLQ) instead minimizes $D(\bp\|\bq)$, our measure of quantization distortion. For both quantizers, pairwise codeword comparisons define half-spaces, so their cells are convex polytopes. We use the cell inradii to derive quantization stability bounds.

For ELQ, we derive bounds on quantization stability (Theorem~\ref{thm:elq-smooth}) and KL distortion (Section~\ref{sec:elq-kl-bound}). For KLQ, we show that each cell can be determined by comparisons with lattice-adjacent codewords (Lemma~\ref{lem:klq-cell-local}), derive a quantization stability bound from the cell inradius (Theorem~\ref{thm:klq-smooth}), and obtain the exact worst-case KL distortion $\ln(K/(K-m+1))$ for the symmetric codebook (Theorem~\ref{thm:klq-kl}). 
We also give a heuristic choice $\alpha^\star$ of the minimum codeword probability (Section~\ref{sec:elq-alpha-heuristic}) and compare it numerically with Lloyd optimization (Section~\ref{sec:klq-lloyd}). Table~\ref{tab:kl_comparison} compares our KL-distortion results with prior work. For ELQ, with $n=km$ and $\beta=\alpha/\delta$ in Reznik's notation, fixing $m$, $\eta$, and $\beta$ gives $\delta=\Theta(n^{-1})$, so our bound has the same $O(n^{-2})$ rate and is nonasymptotic. For KLQ, the worst-case distortion is attained at a simplex vertex. Neither of these prior works studies quantization stability, while we give stability bounds for both ELQ and KLQ.

In simulations, we build canonical Huffman codes from distributions quantized by ELQ, KLQ, and the method of Adler et al.~\cite{adler2023efficient}, and compare them with canonical Huffman coding without quantization. We measure codebook stability under random deletions, KL distortion, and encoded length (Section~\ref{sec:sim}).

\end{vlong}

\section{Preliminaries}
\label{sec:prelim}

\begin{vlong}
Let $m\ge 2$ be an integer. The $(m-1)$-simplex is $\Delta_{m-1}=\{\bp\in\mathbb{R}^m_{\ge 0}:\sum_i p_i=1\}$, with interior $\Delta^\circ_{m-1}$ and tangent subspace $H_0=\{\bx:\sum_i x_i=0\}$.
The $A_{m-1}$ root lattice $\{\bz\in\mathbb{Z}^m:\sum_i z_i=0\}$ has minimal vectors $\be_i-\be_j$ of length $\sqrt{2}$. Figure~\ref{fig:simplex-a2} shows the $m=3$ case.

A \emph{type} of length $K\ge m$ is a distribution on $\{1,\ldots,m\}$ with entries in $\tfrac{1}{K}\mathbb{Z}$. Let $\Sint_K^m=\{\bp\in\Delta^\circ_{m-1}:Kp_i\in\mathbb{Z}_{\ge 1}\ \forall i\}$ denote the \emph{interior types}, and let $\hat{\mathbb{S}}_K^m=\{\bp\in\Delta_{m-1}:Kp_i\in\mathbb{Z}_{\ge 0}\ \forall i,\ \min_i p_i=0\}$ denote the \emph{boundary types}.

For $\bp,\bq\in\Delta_{m-1}$, the total variation distance is
$\mathrm{TV}(\bp,\bq)=\frac{1}{2}\|\bp-\bq\|_1$.
When $q_i>0$ for every $i$ with $p_i>0$, the KL divergence is
$D(\bp\|\bq)=\sum_{i:p_i>0}p_i\ln\frac{p_i}{q_i}$.
We will use the bound
\begin{equation}
\label{eq:kl-from-ratio}
D(\bp\|\bq)\le\sum_{i:p_i>0}p_i\ln \Bigl(\max_{j:p_j>0}\frac{p_j}{q_j}\Bigr)=\ln \max_{i:p_i>0}\frac{p_i}{q_i}.
\end{equation}

A \emph{quantizer} on $\Delta_{m-1}$ has a finite \emph{codebook} $\mathcal{C}\subset\Delta_{m-1}$ and cells $\{V(\bq)\}_{\bq\in\mathcal{C}}$ that partition $\Delta_{m-1}$. The map $\phi:\Delta_{m-1}\to\mathcal{C}$ sends $\bp$ to the codeword $\bq$ whose cell contains $\bp$.

For a perturbation size $\varepsilon$, we define \emph{quantization stability} as
\begin{equation}
S(\varepsilon)=\Pr\bigl(\phi(\bp)=\phi(\bp')\bigr),
\end{equation}
where $\bp\sim\mathrm{Unif}(\Delta_{m-1})$ and, given $\bp$, $\bp'$ is drawn uniformly from the TV-ball of radius $\varepsilon$ around $\bp$. Let $\mathrm{CH}(\cdot)$ denote the canonical Huffman codebook with fixed tie-breaking. We define \emph{codebook stability} under the same perturbation model as $S_{\mathrm{cb}}^\phi(\varepsilon) =\Pr[\mathrm{CH}(\phi(\bp))=\mathrm{CH}(\phi(\bp'))]$. Since equal representatives give the same codebook, $S_{\mathrm{cb}}^\phi(\varepsilon)\ge S(\varepsilon)$.

Define $S_{\mathrm{margin}}(\varepsilon)$ as the probability that a uniformly random $\bp\in\Delta_{m-1}$ is at Euclidean distance at least $2\varepsilon$ from the boundary of its cell. All Euclidean distances to cell boundaries are taken within the affine hull of $\Delta_{m-1}$. Since $\|\bp-\bp'\|_2\le 2\mathrm{TV}(\bp,\bp') \le 2\varepsilon$, such a point $\bp$ remains in the same cell after the perturbation. Hence \begin{equation}
\label{eq:l2-proxy}
S(\varepsilon)\ge S_{\mathrm{margin}}(\varepsilon).
\end{equation}
Theorems~\ref{thm:elq-smooth} and~\ref{thm:klq-smooth} give lower bounds on $S_{\mathrm{margin}}(\varepsilon)$ for ELQ and KLQ, which also bound quantization and codebook stability from below. Section~\ref{sec:sim} evaluates codebook stability under random deletions.

\begin{figure}[t]
\centering
\begin{subfigure}[b]{0.38\linewidth}
\centering
\includegraphics[width=\linewidth]{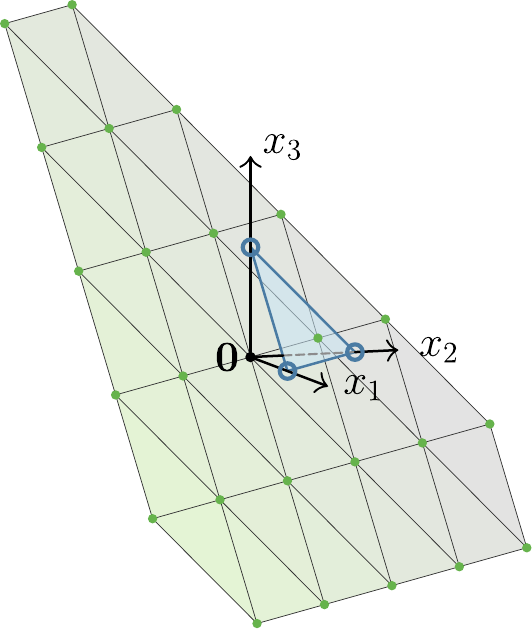}
\caption{}\label{fig:simplex-a2}
\end{subfigure}\hfill
\begin{subfigure}[b]{0.4\linewidth}
\centering
\includegraphics[width=\linewidth]{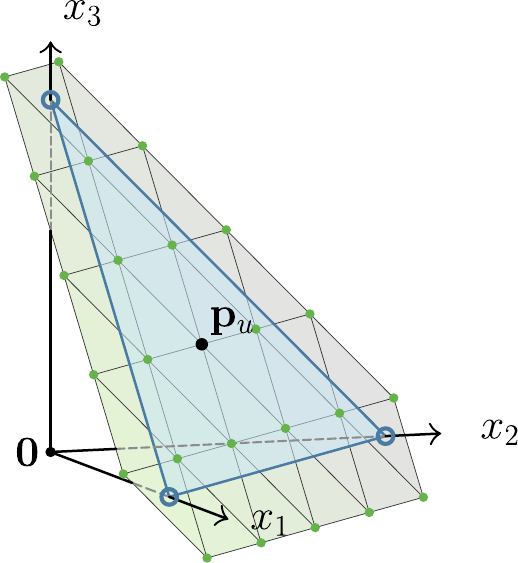}
\caption{}\label{fig:codebook}
\end{subfigure}
\caption{(\subref{fig:simplex-a2}) The 2-simplex $\Delta_2$ (corners shown as open rings, intersecting each axis at $1$) and the $A_2$ root lattice (filled circles) in the parallel zero-sum plane through the origin. (\subref{fig:codebook}) The codebook $\mathcal{C}_\delta$ (\Cref{sec:codebook}) and surrounding lattice $\mathcal{L}_\delta=\pu+\delta A_{m-1}$, centered at the labeled simplex centroid $\pu=(1/3,1/3,1/3)$, for $m=3$, $k=1$, $\delta=1/4$.}
\label{fig:simplex-codebook}
\end{figure}
\end{vlong}
\section{Codebook}
\label{sec:codebook}

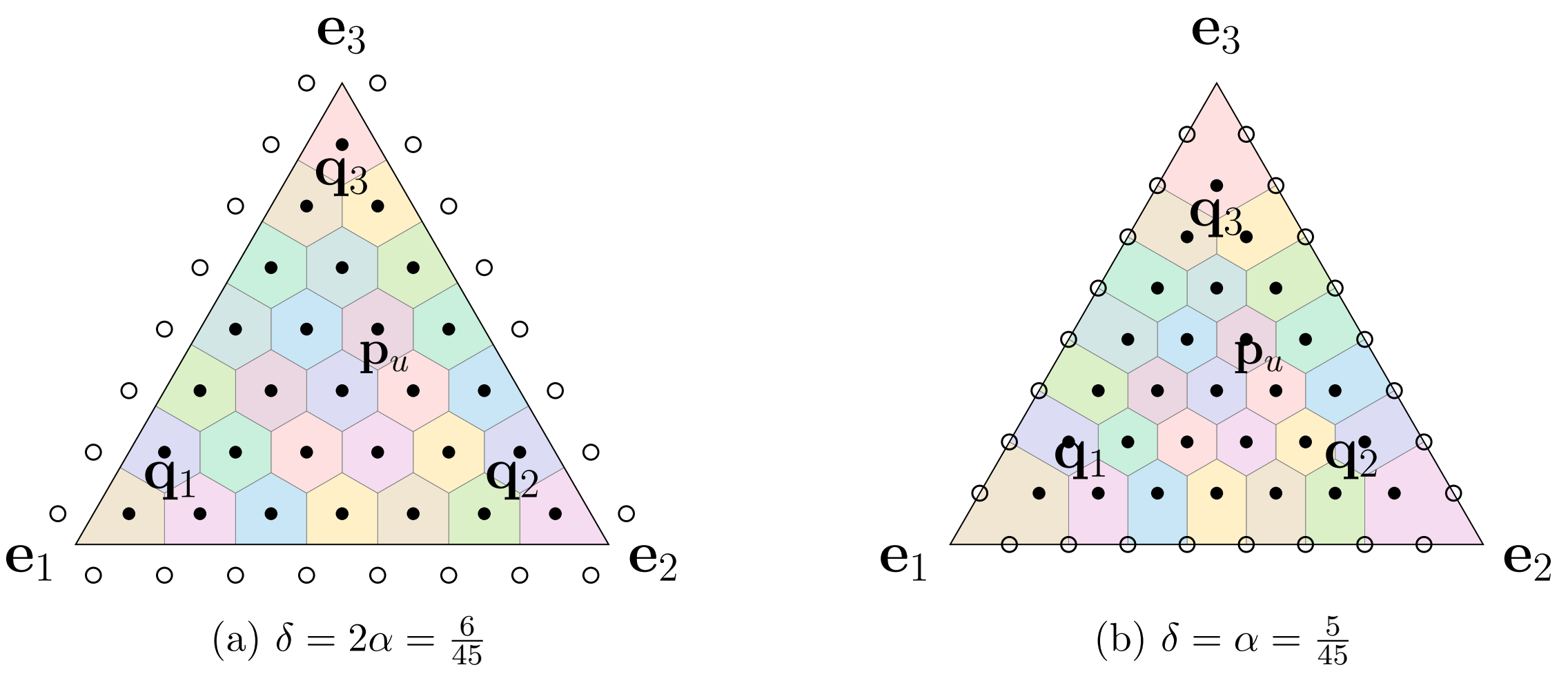
\begin{figure}[!t]
\centering
\begin{subfigure}[!t]{0.45\linewidth}
\centering
\resizebox{\linewidth}{!}{\begin{tikzpicture}[scale=5.5,every node/.append style={font=\Large}]
  \fill[color={rgb,255:red,255;green,224;blue,224}, draw=black!50, line width=0.15pt] (0.50000,0.86603) -- (0.41667,0.72169) -- (0.50000,0.67358) -- (0.58333,0.72169) -- cycle;
  \fill[color={rgb,255:red,255;green,240;blue,200}, draw=black!50, line width=0.15pt] (0.65000,0.60622) -- (0.58333,0.72169) -- (0.50000,0.67358) -- (0.50000,0.59660) -- (0.56667,0.55811) -- cycle;
  \fill[color={rgb,255:red,220;green,240;blue,200}, draw=black!50, line width=0.15pt] (0.56667,0.55811) -- (0.56667,0.48113) -- (0.63333,0.44264) -- (0.71667,0.49075) -- (0.65000,0.60622) -- cycle;
  \fill[color={rgb,255:red,200;green,240;blue,220}, draw=black!50, line width=0.15pt] (0.63333,0.44264) -- (0.63333,0.36566) -- (0.70000,0.32717) -- (0.78333,0.37528) -- (0.71667,0.49075) -- cycle;
  \fill[color={rgb,255:red,200;green,230;blue,245}, draw=black!50, line width=0.15pt] (0.85000,0.25981) -- (0.78333,0.37528) -- (0.70000,0.32717) -- (0.70000,0.25019) -- (0.76667,0.21170) -- cycle;
  \fill[color={rgb,255:red,220;green,220;blue,245}, draw=black!50, line width=0.15pt] (0.91667,0.14434) -- (0.85000,0.25981) -- (0.76667,0.21170) -- (0.76667,0.13472) -- (0.83333,0.09623) -- cycle;
  \fill[color={rgb,255:red,245;green,220;blue,240}, draw=black!50, line width=0.15pt] (0.83333,0.00000) -- (1.00000,0.00000) -- (0.91667,0.14434) -- (0.83333,0.09623) -- cycle;
  \fill[color={rgb,255:red,240;green,230;blue,210}, draw=black!50, line width=0.15pt] (0.50000,0.59660) -- (0.50000,0.67358) -- (0.41667,0.72169) -- (0.35000,0.60622) -- (0.43333,0.55811) -- cycle;
  \fill[color={rgb,255:red,210;green,230;blue,230}, draw=black!50, line width=0.15pt] (0.56667,0.48113) -- (0.56667,0.55811) -- (0.50000,0.59660) -- (0.43333,0.55811) -- (0.43333,0.48113) -- (0.50000,0.44264) -- cycle;
  \fill[color={rgb,255:red,235;green,215;blue,225}, draw=black!50, line width=0.15pt] (0.50000,0.44264) -- (0.50000,0.36566) -- (0.56667,0.32717) -- (0.63333,0.36566) -- (0.63333,0.44264) -- (0.56667,0.48113) -- cycle;
  \fill[color={rgb,255:red,255;green,224;blue,224}, draw=black!50, line width=0.15pt] (0.70000,0.25019) -- (0.70000,0.32717) -- (0.63333,0.36566) -- (0.56667,0.32717) -- (0.56667,0.25019) -- (0.63333,0.21170) -- cycle;
  \fill[color={rgb,255:red,255;green,240;blue,200}, draw=black!50, line width=0.15pt] (0.76667,0.13472) -- (0.76667,0.21170) -- (0.70000,0.25019) -- (0.63333,0.21170) -- (0.63333,0.13472) -- (0.70000,0.09623) -- cycle;
  \fill[color={rgb,255:red,220;green,240;blue,200}, draw=black!50, line width=0.15pt] (0.70000,0.09623) -- (0.70000,0.00000) -- (0.83333,-0.00000) -- (0.83333,0.09623) -- (0.76667,0.13472) -- cycle;
  \fill[color={rgb,255:red,200;green,240;blue,220}, draw=black!50, line width=0.15pt] (0.43333,0.55811) -- (0.35000,0.60622) -- (0.28333,0.49075) -- (0.36667,0.44264) -- (0.43333,0.48113) -- cycle;
  \fill[color={rgb,255:red,200;green,230;blue,245}, draw=black!50, line width=0.15pt] (0.50000,0.36566) -- (0.50000,0.44264) -- (0.43333,0.48113) -- (0.36667,0.44264) -- (0.36667,0.36566) -- (0.43333,0.32717) -- cycle;
  \fill[color={rgb,255:red,220;green,220;blue,245}, draw=black!50, line width=0.15pt] (0.43333,0.32717) -- (0.43333,0.25019) -- (0.50000,0.21170) -- (0.56667,0.25019) -- (0.56667,0.32717) -- (0.50000,0.36566) -- cycle;
  \fill[color={rgb,255:red,245;green,220;blue,240}, draw=black!50, line width=0.15pt] (0.50000,0.21170) -- (0.50000,0.13472) -- (0.56667,0.09623) -- (0.63333,0.13472) -- (0.63333,0.21170) -- (0.56667,0.25019) -- cycle;
  \fill[color={rgb,255:red,240;green,230;blue,210}, draw=black!50, line width=0.15pt] (0.56667,0.09623) -- (0.56667,-0.00000) -- (0.70000,0.00000) -- (0.70000,0.09623) -- (0.63333,0.13472) -- cycle;
  \fill[color={rgb,255:red,210;green,230;blue,230}, draw=black!50, line width=0.15pt] (0.36667,0.36566) -- (0.36667,0.44264) -- (0.28333,0.49075) -- (0.21667,0.37528) -- (0.30000,0.32717) -- cycle;
  \fill[color={rgb,255:red,235;green,215;blue,225}, draw=black!50, line width=0.15pt] (0.30000,0.32717) -- (0.30000,0.25019) -- (0.36667,0.21170) -- (0.43333,0.25019) -- (0.43333,0.32717) -- (0.36667,0.36566) -- cycle;
  \fill[color={rgb,255:red,255;green,224;blue,224}, draw=black!50, line width=0.15pt] (0.50000,0.13472) -- (0.50000,0.21170) -- (0.43333,0.25019) -- (0.36667,0.21170) -- (0.36667,0.13472) -- (0.43333,0.09623) -- cycle;
  \fill[color={rgb,255:red,255;green,240;blue,200}, draw=black!50, line width=0.15pt] (0.43333,0.09623) -- (0.43333,0.00000) -- (0.56667,0.00000) -- (0.56667,0.09623) -- (0.50000,0.13472) -- cycle;
  \fill[color={rgb,255:red,220;green,240;blue,200}, draw=black!50, line width=0.15pt] (0.30000,0.32717) -- (0.21667,0.37528) -- (0.15000,0.25981) -- (0.23333,0.21170) -- (0.30000,0.25019) -- cycle;
  \fill[color={rgb,255:red,200;green,240;blue,220}, draw=black!50, line width=0.15pt] (0.23333,0.21170) -- (0.23333,0.13472) -- (0.30000,0.09623) -- (0.36667,0.13472) -- (0.36667,0.21170) -- (0.30000,0.25019) -- cycle;
  \fill[color={rgb,255:red,200;green,230;blue,245}, draw=black!50, line width=0.15pt] (0.43333,0.00000) -- (0.43333,0.09623) -- (0.36667,0.13472) -- (0.30000,0.09623) -- (0.30000,0.00000) -- cycle;
  \fill[color={rgb,255:red,220;green,220;blue,245}, draw=black!50, line width=0.15pt] (0.23333,0.21170) -- (0.15000,0.25981) -- (0.08333,0.14434) -- (0.16667,0.09623) -- (0.23333,0.13472) -- cycle;
  \fill[color={rgb,255:red,245;green,220;blue,240}, draw=black!50, line width=0.15pt] (0.30000,0.00000) -- (0.30000,0.09623) -- (0.23333,0.13472) -- (0.16667,0.09623) -- (0.16667,0.00000) -- cycle;
  \fill[color={rgb,255:red,240;green,230;blue,210}, draw=black!50, line width=0.15pt] (0.16667,0.09623) -- (0.08333,0.14434) -- (0.00000,0.00000) -- (0.16667,0.00000) -- cycle;

  \draw[black, line width=0.4pt] (0.00000,0.00000) -- (1.00000,0.00000) -- (0.50000,0.86603) -- cycle;

  \draw[black, line width=0.6pt] (0.56667,0.86603) circle (0.014);
  \draw[black, line width=0.6pt] (0.63333,0.75056) circle (0.014);
  \draw[black, line width=0.6pt] (0.70000,0.63509) circle (0.014);
  \draw[black, line width=0.6pt] (0.76667,0.51962) circle (0.014);
  \draw[black, line width=0.6pt] (0.83333,0.40415) circle (0.014);
  \draw[black, line width=0.6pt] (0.90000,0.28868) circle (0.014);
  \draw[black, line width=0.6pt] (0.96667,0.17321) circle (0.014);
  \draw[black, line width=0.6pt] (1.03333,0.05774) circle (0.014);
  \draw[black, line width=0.6pt] (0.43333,0.86603) circle (0.014);
  \draw[black, line width=0.6pt] (0.96667,-0.05774) circle (0.014);
  \draw[black, line width=0.6pt] (0.36667,0.75056) circle (0.014);
  \draw[black, line width=0.6pt] (0.83333,-0.05774) circle (0.014);
  \draw[black, line width=0.6pt] (0.30000,0.63509) circle (0.014);
  \draw[black, line width=0.6pt] (0.70000,-0.05774) circle (0.014);
  \draw[black, line width=0.6pt] (0.23333,0.51962) circle (0.014);
  \draw[black, line width=0.6pt] (0.56667,-0.05774) circle (0.014);
  \draw[black, line width=0.6pt] (0.16667,0.40415) circle (0.014);
  \draw[black, line width=0.6pt] (0.43333,-0.05774) circle (0.014);
  \draw[black, line width=0.6pt] (0.10000,0.28868) circle (0.014);
  \draw[black, line width=0.6pt] (0.30000,-0.05774) circle (0.014);
  \draw[black, line width=0.6pt] (0.03333,0.17321) circle (0.014);
  \draw[black, line width=0.6pt] (0.16667,-0.05774) circle (0.014);
  \draw[black, line width=0.6pt] (-0.03333,0.05774) circle (0.014);
  \draw[black, line width=0.6pt] (0.03333,-0.05774) circle (0.014);

  \fill[black] (0.50000,0.75056) circle (0.012);
  \fill[black] (0.56667,0.63509) circle (0.012);
  \fill[black] (0.63333,0.51962) circle (0.012);
  \fill[black] (0.70000,0.40415) circle (0.012);
  \fill[black] (0.76667,0.28868) circle (0.012);
  \fill[black] (0.83333,0.17321) circle (0.012);
  \fill[black] (0.90000,0.05774) circle (0.012);
  \fill[black] (0.43333,0.63509) circle (0.012);
  \fill[black] (0.50000,0.51962) circle (0.012);
  \fill[black] (0.56667,0.40415) circle (0.012);
  \fill[black] (0.63333,0.28868) circle (0.012);
  \fill[black] (0.70000,0.17321) circle (0.012);
  \fill[black] (0.76667,0.05774) circle (0.012);
  \fill[black] (0.36667,0.51962) circle (0.012);
  \fill[black] (0.43333,0.40415) circle (0.012);
  \fill[black] (0.50000,0.28868) circle (0.012);
  \fill[black] (0.56667,0.17321) circle (0.012);
  \fill[black] (0.63333,0.05774) circle (0.012);
  \fill[black] (0.30000,0.40415) circle (0.012);
  \fill[black] (0.36667,0.28868) circle (0.012);
  \fill[black] (0.43333,0.17321) circle (0.012);
  \fill[black] (0.50000,0.05774) circle (0.012);
  \fill[black] (0.23333,0.28868) circle (0.012);
  \fill[black] (0.30000,0.17321) circle (0.012);
  \fill[black] (0.36667,0.05774) circle (0.012);
  \fill[black] (0.16667,0.17321) circle (0.012);
  \fill[black] (0.23333,0.05774) circle (0.012);
  \fill[black] (0.10000,0.05774) circle (0.012);

  \node[anchor=center] at (-0.08500,-0.03500) {$\mathbf{e}_1$};
  \node[anchor=center] at (1.08500,-0.03500) {$\mathbf{e}_2$};
  \node[anchor=center] at (0.50000,0.95500) {$\mathbf{e}_3$};

  \node[font=\large, above right] at (0.50500,0.29500) {$\pu$};

  \node[above right] at (0.10000,0.05774) {$\bq_1$};
  \node[above left] at (0.90000,0.05774) {$\bq_2$};
  \node[below] at (0.50000,0.75056) {$\bq_3$};
\end{tikzpicture}}
\caption{$\delta=2\alpha=\tfrac{6}{45}$}
\label{fig:cells-elq-hex}
\end{subfigure}\hfill
\begin{subfigure}[!t]{0.45\linewidth}
\centering
\resizebox{\linewidth}{!}{\begin{tikzpicture}[scale=5.5,every node/.append style={font=\Large}]
  \fill[color={rgb,255:red,255;green,224;blue,224}, draw=black!50, line width=0.15pt] (0.50000,0.86603) -- (0.38889,0.67358) -- (0.50000,0.60943) -- (0.61111,0.67358) -- cycle;
  \fill[color={rgb,255:red,255;green,240;blue,200}, draw=black!50, line width=0.15pt] (0.50000,0.60943) -- (0.50000,0.54528) -- (0.55556,0.51320) -- (0.66667,0.57735) -- (0.61111,0.67358) -- cycle;
  \fill[color={rgb,255:red,220;green,240;blue,200}, draw=black!50, line width=0.15pt] (0.55556,0.51320) -- (0.55556,0.44905) -- (0.61111,0.41698) -- (0.72222,0.48113) -- (0.66667,0.57735) -- cycle;
  \fill[color={rgb,255:red,200;green,240;blue,220}, draw=black!50, line width=0.15pt] (0.77778,0.38490) -- (0.72222,0.48113) -- (0.61111,0.41698) -- (0.61111,0.35283) -- (0.66667,0.32075) -- cycle;
  \fill[color={rgb,255:red,200;green,230;blue,245}, draw=black!50, line width=0.15pt] (0.66667,0.32075) -- (0.66667,0.25660) -- (0.72222,0.22453) -- (0.83333,0.28868) -- (0.77778,0.38490) -- cycle;
  \fill[color={rgb,255:red,220;green,220;blue,245}, draw=black!50, line width=0.15pt] (0.72222,0.22453) -- (0.72222,0.16038) -- (0.77778,0.12830) -- (0.88889,0.19245) -- (0.83333,0.28868) -- cycle;
  \fill[color={rgb,255:red,245;green,220;blue,240}, draw=black!50, line width=0.15pt] (1.00000,0.00000) -- (0.88889,0.19245) -- (0.77778,0.12830) -- (0.77778,0.00000) -- cycle;
  \fill[color={rgb,255:red,240;green,230;blue,210}, draw=black!50, line width=0.15pt] (0.50000,0.60943) -- (0.38889,0.67358) -- (0.33333,0.57735) -- (0.44444,0.51320) -- (0.50000,0.54528) -- cycle;
  \fill[color={rgb,255:red,210;green,230;blue,230}, draw=black!50, line width=0.15pt] (0.44444,0.51320) -- (0.44444,0.44905) -- (0.50000,0.41698) -- (0.55556,0.44905) -- (0.55556,0.51320) -- (0.50000,0.54528) -- cycle;
  \fill[color={rgb,255:red,235;green,215;blue,225}, draw=black!50, line width=0.15pt] (0.50000,0.41698) -- (0.50000,0.35283) -- (0.55556,0.32075) -- (0.61111,0.35283) -- (0.61111,0.41698) -- (0.55556,0.44905) -- cycle;
  \fill[color={rgb,255:red,255;green,224;blue,224}, draw=black!50, line width=0.15pt] (0.55556,0.32075) -- (0.55556,0.25660) -- (0.61111,0.22453) -- (0.66667,0.25660) -- (0.66667,0.32075) -- (0.61111,0.35283) -- cycle;
  \fill[color={rgb,255:red,255;green,240;blue,200}, draw=black!50, line width=0.15pt] (0.72222,0.16038) -- (0.72222,0.22453) -- (0.66667,0.25660) -- (0.61111,0.22453) -- (0.61111,0.16038) -- (0.66667,0.12830) -- cycle;
  \fill[color={rgb,255:red,220;green,240;blue,200}, draw=black!50, line width=0.15pt] (0.66667,0.12830) -- (0.66667,0.00000) -- (0.77778,0.00000) -- (0.77778,0.12830) -- (0.72222,0.16038) -- cycle;
  \fill[color={rgb,255:red,200;green,240;blue,220}, draw=black!50, line width=0.15pt] (0.44444,0.44905) -- (0.44444,0.51320) -- (0.33333,0.57735) -- (0.27778,0.48113) -- (0.38889,0.41698) -- cycle;
  \fill[color={rgb,255:red,200;green,230;blue,245}, draw=black!50, line width=0.15pt] (0.50000,0.35283) -- (0.50000,0.41698) -- (0.44444,0.44905) -- (0.38889,0.41698) -- (0.38889,0.35283) -- (0.44444,0.32075) -- cycle;
  \fill[color={rgb,255:red,220;green,220;blue,245}, draw=black!50, line width=0.15pt] (0.44444,0.32075) -- (0.44444,0.25660) -- (0.50000,0.22453) -- (0.55556,0.25660) -- (0.55556,0.32075) -- (0.50000,0.35283) -- cycle;
  \fill[color={rgb,255:red,245;green,220;blue,240}, draw=black!50, line width=0.15pt] (0.50000,0.22453) -- (0.50000,0.16038) -- (0.55556,0.12830) -- (0.61111,0.16038) -- (0.61111,0.22453) -- (0.55556,0.25660) -- cycle;
  \fill[color={rgb,255:red,240;green,230;blue,210}, draw=black!50, line width=0.15pt] (0.66667,0.00000) -- (0.66667,0.12830) -- (0.61111,0.16038) -- (0.55556,0.12830) -- (0.55556,0.00000) -- cycle;
  \fill[color={rgb,255:red,210;green,230;blue,230}, draw=black!50, line width=0.15pt] (0.38889,0.41698) -- (0.27778,0.48113) -- (0.22222,0.38490) -- (0.33333,0.32075) -- (0.38889,0.35283) -- cycle;
  \fill[color={rgb,255:red,235;green,215;blue,225}, draw=black!50, line width=0.15pt] (0.44444,0.25660) -- (0.44444,0.32075) -- (0.38889,0.35283) -- (0.33333,0.32075) -- (0.33333,0.25660) -- (0.38889,0.22453) -- cycle;
  \fill[color={rgb,255:red,255;green,224;blue,224}, draw=black!50, line width=0.15pt] (0.50000,0.16038) -- (0.50000,0.22453) -- (0.44444,0.25660) -- (0.38889,0.22453) -- (0.38889,0.16038) -- (0.44444,0.12830) -- cycle;
  \fill[color={rgb,255:red,255;green,240;blue,200}, draw=black!50, line width=0.15pt] (0.44444,0.12830) -- (0.44444,0.00000) -- (0.55556,0.00000) -- (0.55556,0.12830) -- (0.50000,0.16038) -- cycle;
  \fill[color={rgb,255:red,220;green,240;blue,200}, draw=black!50, line width=0.15pt] (0.33333,0.25660) -- (0.33333,0.32075) -- (0.22222,0.38490) -- (0.16667,0.28868) -- (0.27778,0.22453) -- cycle;
  \fill[color={rgb,255:red,200;green,240;blue,220}, draw=black!50, line width=0.15pt] (0.27778,0.22453) -- (0.27778,0.16038) -- (0.33333,0.12830) -- (0.38889,0.16038) -- (0.38889,0.22453) -- (0.33333,0.25660) -- cycle;
  \fill[color={rgb,255:red,200;green,230;blue,245}, draw=black!50, line width=0.15pt] (0.33333,0.12830) -- (0.33333,0.00000) -- (0.44444,-0.00000) -- (0.44444,0.12830) -- (0.38889,0.16038) -- cycle;
  \fill[color={rgb,255:red,220;green,220;blue,245}, draw=black!50, line width=0.15pt] (0.27778,0.22453) -- (0.16667,0.28868) -- (0.11111,0.19245) -- (0.22222,0.12830) -- (0.27778,0.16038) -- cycle;
  \fill[color={rgb,255:red,245;green,220;blue,240}, draw=black!50, line width=0.15pt] (0.33333,0.00000) -- (0.33333,0.12830) -- (0.27778,0.16038) -- (0.22222,0.12830) -- (0.22222,-0.00000) -- cycle;
  \fill[color={rgb,255:red,240;green,230;blue,210}, draw=black!50, line width=0.15pt] (0.22222,-0.00000) -- (0.22222,0.12830) -- (0.11111,0.19245) -- (0.00000,0.00000) -- cycle;

  \draw[black, line width=0.4pt] (0.00000,0.00000) -- (1.00000,0.00000) -- (0.50000,0.86603) -- cycle;

  \draw[black, line width=0.6pt] (0.55556,0.76980) circle (0.014);
  \draw[black, line width=0.6pt] (0.61111,0.67358) circle (0.014);
  \draw[black, line width=0.6pt] (0.66667,0.57735) circle (0.014);
  \draw[black, line width=0.6pt] (0.72222,0.48113) circle (0.014);
  \draw[black, line width=0.6pt] (0.77778,0.38490) circle (0.014);
  \draw[black, line width=0.6pt] (0.83333,0.28868) circle (0.014);
  \draw[black, line width=0.6pt] (0.88889,0.19245) circle (0.014);
  \draw[black, line width=0.6pt] (0.94444,0.09623) circle (0.014);
  \draw[black, line width=0.6pt] (0.44444,0.76980) circle (0.014);
  \draw[black, line width=0.6pt] (0.88889,0.00000) circle (0.014);
  \draw[black, line width=0.6pt] (0.38889,0.67358) circle (0.014);
  \draw[black, line width=0.6pt] (0.77778,0.00000) circle (0.014);
  \draw[black, line width=0.6pt] (0.33333,0.57735) circle (0.014);
  \draw[black, line width=0.6pt] (0.66667,0.00000) circle (0.014);
  \draw[black, line width=0.6pt] (0.27778,0.48113) circle (0.014);
  \draw[black, line width=0.6pt] (0.55556,0.00000) circle (0.014);
  \draw[black, line width=0.6pt] (0.22222,0.38490) circle (0.014);
  \draw[black, line width=0.6pt] (0.44444,0.00000) circle (0.014);
  \draw[black, line width=0.6pt] (0.16667,0.28868) circle (0.014);
  \draw[black, line width=0.6pt] (0.33333,0.00000) circle (0.014);
  \draw[black, line width=0.6pt] (0.11111,0.19245) circle (0.014);
  \draw[black, line width=0.6pt] (0.22222,0.00000) circle (0.014);
  \draw[black, line width=0.6pt] (0.05556,0.09623) circle (0.014);
  \draw[black, line width=0.6pt] (0.11111,0.00000) circle (0.014);

  \fill[black] (0.50000,0.67358) circle (0.012);
  \fill[black] (0.55556,0.57735) circle (0.012);
  \fill[black] (0.61111,0.48113) circle (0.012);
  \fill[black] (0.66667,0.38490) circle (0.012);
  \fill[black] (0.72222,0.28868) circle (0.012);
  \fill[black] (0.77778,0.19245) circle (0.012);
  \fill[black] (0.83333,0.09623) circle (0.012);
  \fill[black] (0.44444,0.57735) circle (0.012);
  \fill[black] (0.50000,0.48113) circle (0.012);
  \fill[black] (0.55556,0.38490) circle (0.012);
  \fill[black] (0.61111,0.28868) circle (0.012);
  \fill[black] (0.66667,0.19245) circle (0.012);
  \fill[black] (0.72222,0.09623) circle (0.012);
  \fill[black] (0.38889,0.48113) circle (0.012);
  \fill[black] (0.44444,0.38490) circle (0.012);
  \fill[black] (0.50000,0.28868) circle (0.012);
  \fill[black] (0.55556,0.19245) circle (0.012);
  \fill[black] (0.61111,0.09623) circle (0.012);
  \fill[black] (0.33333,0.38490) circle (0.012);
  \fill[black] (0.38889,0.28868) circle (0.012);
  \fill[black] (0.44444,0.19245) circle (0.012);
  \fill[black] (0.50000,0.09623) circle (0.012);
  \fill[black] (0.27778,0.28868) circle (0.012);
  \fill[black] (0.33333,0.19245) circle (0.012);
  \fill[black] (0.38889,0.09623) circle (0.012);
  \fill[black] (0.22222,0.19245) circle (0.012);
  \fill[black] (0.27778,0.09623) circle (0.012);
  \fill[black] (0.16667,0.09623) circle (0.012);

  \node[anchor=center] at (-0.08500,-0.03500) {$\mathbf{e}_1$};
  \node[anchor=center] at (1.08500,-0.03500) {$\mathbf{e}_2$};
  \node[anchor=center] at (0.50000,0.95500) {$\mathbf{e}_3$};

  \node[font=\large, above right] at (0.50500,0.29500) {$\pu$};

  \node[above right] at (0.16667,0.09623) {$\bq_1$};
  \node[above left] at (0.83333,0.09623) {$\bq_2$};
  \node[below] at (0.50000,0.67358) {$\bq_3$};
\end{tikzpicture}}
\caption{$\delta=\alpha=\tfrac{5}{45}$}
\label{fig:cells-elq-sym}
\end{subfigure}
\caption{ELQ ($\ell_2$-Voronoi) cell partitions at $m=3$, $k=2$ ($K=9$, 28 codewords). In (\subref{fig:cells-elq-hex}) $\alpha=\delta/2$, and lattice neighbors obtained by decreasing a minimum coordinate lie outside the simplex. In (\subref{fig:cells-elq-sym}), $\alpha=\delta$, and these neighbors lie on the simplex boundary.}
\label{fig:cells-elq}
\end{figure}

\begin{vlong}
Both quantizers studied in this paper use the same lattice-based codebook, namely a shifted and scaled version $\mathcal L_\delta$ of the $A_{m-1}$ lattice, restricted to the interior of the probability simplex $\Delta_{m-1}$. Fix $\delta\in(0,1/m)$ and let $\pu=\bone/m$ be the uniform distribution. Define
\[
\mathcal L_\delta\triangleq\pu+\delta A_{m-1},\qquad
\mathcal{C}_\delta\triangleq\mathcal L_\delta\cap\Delta^\circ_{m-1}.
\]
We exclude codewords on the boundary so that the KL divergence to any codeword is finite. An example for $m=3$ is shown in Figure~\ref{fig:codebook}. For later use, we write this codebook in terms of $\alpha$, $\delta$, and $k$.

\begin{lemma}
\label{lem:codebook-repr}
For $\delta \in (0, 1/m)$, we have
\begin{equation}
\label{eq:codebook-wform}
\mathcal{C}_\delta=\bigl\{\alpha\bone+\delta\bw:\bw\in\mathbb{Z}_{\ge 0}^m,\ \textstyle\sum_i w_i=km\bigr\},
\end{equation}
where
\begin{equation}
\label{eq:k-alpha}
k=\ceil{\tfrac{1}{m\delta}}-1,\qquad
\alpha=\tfrac{1}{m}-k\delta\in(0,\delta].
\end{equation}
Furthermore,
$|\mathcal{C}_\delta|=\binom{km+m-1}{m-1}$.
\end{lemma}

\begin{proof}
Any $\bq\in\mathcal{L}_\delta$ has the form $\bq=\pu+\delta\bz$ with $\bz\in A_{m-1}$. Since $\pu=(\alpha+k\delta)\bone$, setting $\bw=\bz+k\bone$ gives $\bq=\alpha\bone+\delta\bw$ and a bijection between $A_{m-1}$ and $\{\bw\in\mathbb{Z}^m:\sum_i w_i=km\}$. Also, $q_i=\alpha+\delta w_i>0$ implies $w_i>-\alpha/\delta$. The definition of $k$ gives $\alpha\in(0,\delta]$, so $-\alpha/\delta\in[-1,0)$. Thus, over the integers, $q_i>0$ is equivalent to $w_i\ge0$. This gives \eqref{eq:codebook-wform}, and the cardinality follows by stars-and-bars.
\end{proof}

In Reznik's notation, $q_i=(w_i+\beta)/(n+m\beta)$, which matches our parametrization with $n=km$ and $\beta=\alpha/\delta$.

Here, $k$ is the number of codewords strictly between $\pu$ and any simplex vertex $\be_i$. Figure~\ref{fig:cells-elq} shows an example with $m=3$ and $k=2$. The parameter $\alpha$ is the smallest value any coordinate of a codeword can take. The \emph{corner} vectors, which are closest to the simplex vertices $\be_i$, are
\[
\bq_i=(\alpha,\ldots,\alpha,
\underbrace{\alpha+km\delta}_{\text{position }i},
\alpha,\ldots,\alpha),\qquad i=1,\ldots,m,
\]
where $\alpha+km\delta=1-(m-1)\alpha$ since the coordinates sum to~1.

For this codebook, set $K\triangleq m(k+1)$. Then Lemma~\ref{lem:codebook-repr} gives
\begin{equation}
\label{eq:card}
|\mathcal{C}_\delta|=\binom{K-1}{m-1}=|\Sint_K^m|.
\end{equation}
For fixed $k$, the valid range is $\alpha\in(0,1/K]$. The set equality $\mathcal{C}_\delta=\Sint_K^m$ requires the symmetric choice $\alpha=\delta=1/K$, where the lattice spacing equals the boundary offset, as shown in \Cref{fig:cells-elq-sym}. In Reznik's biased-type notation, this choice corresponds to $n=K-m$ and $\beta=1$. In this case, lattice points on the simplex boundary are excluded from the codebook.
\end{vlong}
\section{Euclidean Lattice Quantizer (ELQ)}
\label{sec:elq}

\begin{vlong}
In the previous section, we introduced the lattice-based codebook $\mathcal{C}_\delta$. In this section, we consider the Euclidean quantizer based on this codebook, which quantizes distributions in $\Delta_{m-1}$ to the nearest codeword in $\mathcal{C}_\delta$ using the Euclidean distance. Reznik~\cite{reznik2011} also studied Euclidean nearest-neighbor quantization on type lattices. Here we use the biased interior codebook $\mathcal C_\delta$ and focus on quantization stability under distribution perturbations.
\end{vlong}

\subsection{Voronoi Cells of the Quantizer}

\begin{definition}
\label{def:elq}
The Euclidean Lattice Quantizer is
\[
\phi_{\mathrm{ELQ}}(\bp)\in\arg\min\nolimits_{\bq\in\mathcal{C}_\delta}\|\bp-\bq\|_2,
\]
with ties broken by a fixed deterministic rule. Its cells are
$V_{\mathrm{ELQ}}(\bq)=\{\bp\in\Delta_{m-1}:\|\bp-\bq\|_2\le\|\bp-\bq'\|_2\ \forall\bq'\in\mathcal{C}_\delta\}$.
\end{definition}

Each pairwise Euclidean comparison defines a half-space, so every $V_{\mathrm{ELQ}}(\bq)$ is a convex polytope. Figure~\ref{fig:cells-elq} shows examples of the Voronoi cells for different choices of $\alpha$. The choice $\alpha=\delta$ is easier to analyze.

For the unrestricted lattice $\mathcal{L}_\delta\supset\mathcal{C}_\delta$, the Voronoi cell at $\bq$ has vertices $\bq+\delta\sigma(\bu^{(j)})$, where $\sigma$ ranges over coordinate permutations, $j\in\{1,\ldots,m-1\}$, and
\begin{equation}
\label{rem:lattice-voronoi-vertices}
\bu^{(j)}=(\underbrace{j/m,\ldots,j/m}_{m-j},\underbrace{(j-m)/m,\ldots,(j-m)/m}_{j}),
\end{equation}
with $\|\bu^{(j)}\|_2^2=j(m-j)/m\le m/4$ \cite[\S 21.3.B]{conway1999sphere}. So every $\bp$ in the unrestricted lattice Voronoi cell of $\bq$ satisfies $\|\bp-\bq\|_2\le \tfrac{1}{2}\delta\sqrt{m}$.

\subsection{Inradius and stability}
\label{sec:elq-smooth}

We first lower-bound the ELQ cell inradius and then use this to lower-bound $S_{\mathrm{margin}}(\varepsilon)$ and hence the quantization stability $S(\varepsilon)$.

\begin{proposition}
\label{prop:elq-inradius}
For every $\bq\in\mathcal{C}_\delta$, the inradius $\rho_{\mathrm{ELQ}}(\bq)\coloneqq \mathrm{dist}(\bq,\partial V_{\mathrm{ELQ}}(\bq))$ satisfies
\begin{equation*}
\rho_{\mathrm{ELQ}}(\bq)
\ge
\min\left(
\frac{\delta}{\sqrt{2}},
\alpha\sqrt{\frac{m}{m-1}}
\right),
\end{equation*}
with equality whenever $\bq$ has a lattice-adjacent neighbour in $\mathcal{C}_\delta$ and the minimum is attained by the first term (i.e., $\alpha\ge\delta\sqrt{\frac{m-1}{2m}}$).
\end{proposition}

\begin{vlong}
\begin{proof}
By Definition~\ref{def:elq}, the cell $V_{\mathrm{ELQ}}(\bq)$ is the intersection of the bisector half-spaces for all $\bq'\in\mathcal{C}_\delta\setminus\{\bq\}$ and the simplex $\Delta_{m-1}$. The inradius $\rho_{\mathrm{ELQ}}(\bq)$ is the minimum $\ell_2$-distance from $\bq$ to any of these bounding hyperplanes.

For the bisectors, the distance to the perpendicular bisector of $\bq'$ is $\|\bq-\bq'\|_2/2$. A lattice-adjacent neighbour $\bq'=\bq+\delta(\be_a-\be_b)$ gives $\|\bq-\bq'\|_2=\delta\sqrt{2}$, so its bisector is at distance $\delta/\sqrt{2}$. Any non-adjacent neighbour $\bq''$ has $\|\bq-\bq''\|_2\ge2\delta$, so its bisector is at distance at least $\delta$. Thus, the minimum bisector distance is $\delta/\sqrt{2}$ whenever a lattice-adjacent neighbour exists.

For the simplex facets, within the affine hyperplane $H=\{\bp:\sum_j p_j=1\}$, the normal vector to the facet $\{p_i=0\}$ is $\be_i-\frac{1}{m}\bone$. Moving from $\bq$ by $-\frac{q_i}{1-(1/m)}(\be_i-\frac{1}{m}\bone)$ reaches this facet. Its Euclidean length is $q_i\sqrt{\frac{m}{m-1}}\ge \alpha\sqrt{\frac{m}{m-1}}$. Taking the minimum over the bisectors and simplex facets gives $\rho_{\mathrm{ELQ}}(\bq)\ge \min\bigl(\frac{\delta}{\sqrt{2}}, \alpha\sqrt{\frac{m}{m-1}}\bigr)$.
\end{proof}
\end{vlong}

\begin{theorem}[ELQ quantization stability]
\label{thm:elq-smooth}
Let $\rho=\min\bigl(\frac{\delta}{\sqrt{2}}, \alpha\sqrt{\frac{m}{m-1}}\bigr)$. For all $\varepsilon\in[0,\rho/2]$,
\begin{equation}
S(\varepsilon)\ge S_{\mathrm{margin}}(\varepsilon)
\ge
\left(1-\frac{2\varepsilon}{\rho}\right)^{m-1}.
\end{equation}
In particular, if $\alpha\ge\delta\sqrt{\frac{m-1}{2m}}$, then $\rho=\delta/\sqrt{2}$ and $S(\varepsilon)\ge(1-2\sqrt{2}\varepsilon/\delta)^{m-1}$.
\end{theorem}

\begin{vlong}
\begin{proof}
By \eqref{eq:l2-proxy}, $S(\varepsilon)\ge S_{\mathrm{margin}}(\varepsilon)$. The case $\varepsilon=0$ is immediate, so assume $\varepsilon>0$.

Fix a cell $V_{\mathrm{ELQ}}(\bq)$ and set $r=\rho$. By Proposition~\ref{prop:elq-inradius}, $\bq$ is at distance at least $r$ from the boundary of the cell. Consider any convex polytope $P\subset\mathbb{R}^{m-1}$ containing a point $c$ at distance at least $r$ from $\partial P$. Scale $P$ toward $c$ by the factor $1-2\varepsilon/r$:
\[
T(x)=c+\left(1-\frac{2\varepsilon}{r}\right)(x-c).
\]
We show that every point of $T(P)$ is at least $2\varepsilon$ from $\partial P$.

Since $B(c,r)\subseteq P$, set $\lambda=2\varepsilon/r\in(0,1]$. For any $z=c+(1-\lambda)(x-c)\in T(P)$ and any $w$ with $\|w\|\le2\varepsilon$, let $\tilde w=w/\lambda$. Then $\|\tilde w\|\le r$, so $c+\tilde w\in P$, and
\[
z+w=(1-\lambda)x+\lambda(c+\tilde w).
\]
This is a convex combination of two points in $P$, so $z+w\in P$. Hence every point of $T(P)$ is at least $2\varepsilon$ from $\partial P$.

Scaling by $1-2\varepsilon/r$ multiplies the $(m-1)$-dimensional volume by $(1-2\varepsilon/r)^{m-1}$. Therefore,
\[
\mathrm{Vol}\bigl(\{x\in P:
\mathrm{dist}(x,\partial P)\ge2\varepsilon\}\bigr)
\ge
\left(1-\frac{2\varepsilon}{r}\right)^{m-1}
\mathrm{Vol}(P).
\]
Applying this to every ELQ cell with $c=\bq$ gives $S_{\mathrm{margin}}(\varepsilon)\ge (1-2\varepsilon/\rho)^{m-1}$.
\end{proof}
\end{vlong}

\subsection{KL distortion bound}
\label{sec:elq-kl-bound}

We next bound $D(\bp\|\phi_{\mathrm{ELQ}}(\bp))$. We first consider lattice cells that are fully contained in the simplex, and then give a bound that applies to every $\bp\in\Delta_{m-1}$.

\begin{lemma}[KL bound under unclipping]
\label{lem:elq-lattice-cell}
Let $\bq\in\mathcal{C}_\delta$ and let $V_{\mathrm{lat}}(\bq)$ denote its Voronoi cell with respect to the unrestricted lattice $\mathcal{L}_\delta\supset\mathcal{C}_\delta$. If
$V_{\mathrm{lat}}(\bq)\subseteq\Delta_{m-1}$, then
$\min_i q_i\ge\delta(m-1)/m$, and for every
$\bp\in V_{\mathrm{lat}}(\bq)$,
\begin{equation}
D(\bp\|\bq)
\le
\frac{m\delta^2}
{4\max\bigl(\alpha,\delta(m-1)/m\bigr)}.
\end{equation}
\end{lemma}

\begin{vlong}
\begin{proof}
Recall from \Cref{rem:lattice-voronoi-vertices} that the vertices of $V_{\mathrm{lat}}(\bq)$ are $\bq+\delta\sigma(\bu^{(j)})$ with $\|\bu^{(j)}\|_2^2=j(m-j)/m\le m/4$.

The unclipping assumption requires $\bq+\delta\bv\in\Delta_{m-1}$ for every Voronoi vertex $\delta\bv$. The most negative single entry over all such $\bv$ is $-(m-1)/m$, attained at $\bu^{(1)}$; cycling this entry through each coordinate position via permutations forces $q_i\ge\delta(m-1)/m$ for all $i$, hence $\min_i q_i\ge\max(\alpha,\,\delta(m-1)/m)$ when combined with the codebook bound. The inequalities
$D(\bp\|\bq)\le\sum_i (p_i-q_i)^2/q_i\le\|\bp-\bq\|_2^2/\min_i q_i$,
together with the lattice covering bound $\|\bp-\bq\|_2^2\le m\delta^2/4$ stated above, yield the stated result.
\end{proof}
\end{vlong}

\begin{lemma}[Global ELQ bound]
\label{lem:elq-l2-coarse-gen}
For every $\bp\in\Delta_{m-1}$ and $\bq=\phi_{\mathrm{ELQ}}(\bp)$ with any valid $\alpha$ and $\delta$,
\begin{equation}
\|\bp-\bq\|_2 \;\le\; \tfrac{1}{2}\delta\sqrt{m} + \alpha\sqrt{m(m-1)}.
\end{equation}
Consequently, using $\min_i q_i\ge\alpha$,
\begin{equation}
D(\bp\|\bq) \;\le\; \frac{\bigl(\tfrac{1}{2}\delta\sqrt{m} + \alpha\sqrt{m(m-1)}\bigr)^2}{\alpha}.
\end{equation}
\end{lemma}

\begin{vlong}
\begin{proof}
We construct a specific $\tilde\bq\in\mathcal{C}_\delta$ near $\bp$; since ELQ chooses the nearest codeword in $\ell_2$ distance, the same bound then holds for $\phi_{\mathrm{ELQ}}(\bp)$.

\emph{Step 1 (move into the interior).} Set $\bx=(1-m\alpha)\bp+\alpha\bone$. Then $\bx\in\Delta_{m-1}$ with $\min_i x_i\ge\alpha$, and
\[
\|\bx-\bp\|_2 \;=\; \alpha\,\|\bone-m\bp\|_2 \;\le\; \alpha\sqrt{m(m-1)},
\]
since $\|\bone-m\bp\|_2^2$ is convex in $\bp$ and attains its maximum on $\Delta_{m-1}$ at a vertex $\be_j$, where $\|\bone-m\be_j\|_2^2=(m-1)+(m-1)^2=m(m-1)$.

\emph{Step 2 (round to $\mathcal{C}_\delta$).} By the $\bw$-parametrization of the codebook in \eqref{eq:codebook-wform}, every $\tilde\bq\in\mathcal{C}_\delta$ has the form $\tilde\bq=\alpha\bone+\delta\bw$ with $\bw\in\mathbb{Z}_{\ge 0}^m$ and $\sum_i w_i=km$. Define $\by\triangleq(\bx-\alpha\bone)/\delta$; since $x_i\ge\alpha$ for every $i$, $\by\in\mathbb{R}_{\ge 0}^m$ and $\sum_i y_i=(1-m\alpha)/\delta=km$. The $\ell_2$-nearest integer vector with $\sum_i w_i=km$ is obtained by the standard nearest-type rounding rule~\cite{reznik2011}: take the floors of the coordinates of $\by$ and assign the remaining counts to the largest residues. Since $y_i\ge0$ for all $i$, the resulting vector also satisfies $w_i\ge0$. Hence $\tilde\bq\triangleq\alpha\bone+\delta\bw$ belongs to $\mathcal C_\delta$ and is a nearest point of the unrestricted lattice to $\bx$, so $\bx\in V_{\mathrm{lat}}(\tilde\bq)$. The lattice covering bound above gives
\[
\|\bx-\tilde\bq\|_2 \;\le\; \tfrac{1}{2}\delta\sqrt{m}.
\]

\emph{Step 3 (combine).} By the triangle inequality and Steps 1--2,
\[
\|\bp-\tilde\bq\|_2 \;\le\; \|\bp-\bx\|_2+\|\bx-\tilde\bq\|_2 \;\le\; \alpha\sqrt{m(m-1)} + \tfrac{1}{2}\delta\sqrt{m}.
\]
Since $\bq=\phi_{\mathrm{ELQ}}(\bp)$ is the $\ell_2$-nearest point of $\mathcal{C}_\delta$ to $\bp$, the same bound holds with $\tilde\bq$ replaced by $\bq$, which proves the first claim. For the KL inequality,
$D(\bp\|\bq)\le\sum_i (p_i-q_i)^2/q_i\le\|\bp-\bq\|_2^2/\min_i q_i$,
and every codeword in $\mathcal{C}_\delta$ satisfies $\min_i q_i\ge\alpha$. Substituting the $\ell_2$ bound gives the stated KL bound.
\end{proof}
\end{vlong}

Reznik~\cite{reznik2011} derived an asymptotic $O(n^{-2})$ KL-distortion result away from the simplex boundary, with $n=km$ in our notation, and a worse bound near the boundary. For unclipped cells with $\min_i q_i\ge\eta>0$ independent of $n$, the argument in Lemma~\ref{lem:elq-lattice-cell} gives the same $O(n^{-2})$ rate. Lemma~\ref{lem:elq-l2-coarse-gen} gives a separate bound over the full simplex.

\begin{vlong}
\subsection{\texorpdfstring{$\alpha$-selection heuristic}{Alpha-selection heuristic}}
\label{sec:elq-alpha-heuristic}

The symmetric choice $\alpha=\delta=1/K$ simplifies the analysis, but need not minimize the distortion. Reznik suggested $\beta=1/m$ for $L_r$-norm quantization, which corresponds to $\alpha/\delta=1/m$ in our notation~\cite{reznik2011}. Here we choose $\alpha$ using a boundary--interior KL balance. Appendix A gives a local condition under which increasing $\alpha$ reduces the KL distortion away from the boundary. At the symmetric choice, numerical enumeration for $m\in\{2,3,4\}$ and $K\le 20$ shows that the worst-case ELQ distortion is attained at a simplex vertex $\bp=\be_j$. For general $\alpha$, $\phi_{\mathrm{ELQ}}(\be_j)=\bq_j=(\alpha,\ldots,\alpha,1-(m-1)\alpha,\alpha,\ldots,\alpha)$, giving distortion

\begin{equation}
B(\alpha)\triangleq D(\be_j\|\bq_j)=\ln\frac{1}{1-(m-1)\alpha}.
\end{equation}
$B(\alpha)$ is strictly increasing in $\alpha$, so a smaller $\alpha$ gives lower distortion at the simplex vertices. To estimate the interior KL distortion, consider a Voronoi vertex of the corner cell, $\bp^\star=\bq_j+\delta\bu^{(1)}$, with the negative entry of $\bu^{(1)}=(1/m,\ldots,1/m,-(m-1)/m)$ aligned with the large coordinate of $\bq_j$. A second-order expansion in $\delta$ gives
\[
D(\bp^\star\|\bq_j)
\approx
\frac{\delta^2}{2m^2}
\left(
\frac{m-1}{\alpha}
+
\frac{(m-1)^2}{1-(m-1)\alpha}
\right).
\]
In this second-order approximation, for small $\alpha$, the first term dominates, so $D(\bp^\star\|\bq_j)\approx\frac{(m-1)\delta^2}{2m^2\alpha}$. Setting this interior approximation equal to $B(\alpha)\approx(m-1)\alpha$ gives $\delta\approx m\sqrt{2}\,\alpha$. Together with $\delta=(1/m-\alpha)/k$, this gives $\alpha\approx1/[m(1+\sqrt{2}mk)]$. For large $mk$, this simplifies to
\begin{equation}
\label{eq:alpha-star}
\alpha^\star(m,k)\approx\frac{1}{m^2k\sqrt{2}}.
\end{equation}

For $m=3$, this gives $\alpha^\star\approx 1/(9\sqrt{2}\,k)\approx0.0786/k$, which is smaller than the symmetric choice $\alpha=\delta=1/[m(k+1)]$.
\end{vlong}

\section{KL Quantizer (KLQ)}
\label{sec:klq}

Since KL divergence is our measure of quantization distortion, we also consider a quantizer that directly minimizes it. For the symmetric codebook $\alpha=\delta=1/K$ and $p_i>0$ for all $i$, this becomes the $K$-type approximation problem studied by Geiger and Böcherer~\cite{geiger2016greedy}. Their work focuses on finding the KL-optimal type efficiently, while here we use the lattice structure to study the KLQ cells and quantization stability.

\subsection{Definition and convex cells}

\begin{definition}
\label{def:klq}
The KL Quantizer is
\[
\phi_{\mathrm{KL}}(\bp)\in\arg\min\nolimits_{\bq\in\mathcal{C}_\delta}D(\bp\|\bq),
\]
with ties broken by a fixed deterministic rule. For the analysis, we use the closed KL Voronoi regions
$V_{\mathrm{KL}}(\bq)=\{\bp\in\Delta_{m-1}:D(\bp\|\bq)\le D(\bp\|\bq')\
\forall\bq'\in\mathcal C_\delta\}$.
The tie-breaking choice only affects their shared boundaries.
\end{definition}

\begin{vlong}
Although $D(\bp\|\bq)$ is nonlinear in $\bp$, the difference
$D(\bp\|\bq)-D(\bp\|\bq')$ is linear in $\bp$, so the cell boundaries are hyperplanes.

\begin{lemma}
\label{lem:kl-convex}
For $\bq,\bq'\in\mathcal{C}_\delta$, the set
$\{\bp\in\Delta_{m-1}:D(\bp\|\bq)\le D(\bp\|\bq')\}$
is the intersection of $\Delta_{m-1}$ with a half-space.
Thus, every $V_{\mathrm{KL}}(\bq)$ is a convex polytope.
\end{lemma}

\begin{proof}
$D(\bp\|\bq)-D(\bp\|\bq')=\sum_i p_i\ln(q'_i/q_i)$
after the $p_i\ln p_i$ terms cancel. This is linear in $\bp$.
\end{proof}
\end{vlong}

\paragraph{KL cells from local comparisons.}

The KL objective over the integer coordinates is a separable convex allocation problem
of the type studied by Geiger and Böcherer~\cite{geiger2016greedy}.
We use this structure to show that it is enough to compare with
lattice-adjacent codewords when defining a KLQ cell.

By Lemma~\ref{lem:codebook-repr}, every codeword
$\bq\in\mathcal{C}_\delta$ has a unique representation
$\bq=\alpha\bone+\delta\bw$ where
$\bw\in\mathbb{Z}_{\ge 0}^m$ with $\sum_i w_i=km$;
we call $\bw$ the \emph{integer coordinate} of $\bq$,
and identify $\bq$ with $\bw$ when convenient.
A \emph{lattice-adjacent} swap from $\bq$ is the codeword
$\bq'=\bq+\delta(\be_a-\be_b)$ whose integer coordinate is
$\bw'=\bw+\be_a-\be_b$; this is well-defined whenever $w_b\ge 1$
(so that $\bw'\ge\mathbf{0}$).
Write $\mathcal{N}(\bq)\subset\mathcal{C}_\delta$ for the set of such neighbours.

\begin{lemma}[Adjacent comparisons determine the KL cell]
\label{lem:klq-cell-local}
For every $\bq\in\mathcal{C}_\delta$,
\[
V_{\mathrm{KL}}(\bq)
=
\{\bp\in\Delta_{m-1}:
D(\bp\|\bq)\le D(\bp\|\bq')
\ \forall\bq'\in\mathcal{N}(\bq)\}.
\]
\end{lemma}

\begin{vlong}
\begin{proof}
The inclusion $V_{\mathrm{KL}}(\bq)\subseteq\{\ldots\}$ is immediate.
For the reverse, write $\bq=\alpha\bone+\delta\bw$ and define
$g_i(t)=-p_i\ln(\alpha+\delta t)$ and
$f(\bv)=\sum_i g_i(v_i)$.
Minimizing $D(\bp\|\bq)$ over $\mathcal C_\delta$
is equivalent to minimizing $f(\bw)$ over
$\bw\in\mathbb Z_{\ge0}^m$ with $\sum_i w_i=km$.
Each $g_i$ is convex, since
$g_i''(t)=p_i\delta^2/(\alpha+\delta t)^2\ge0$.

Suppose no lattice-adjacent swap decreases $f$.
Then, for every $i,j$ with $w_j\ge1$,
\[
g_i(w_i+1)-g_i(w_i)
\ge
g_j(w_j)-g_j(w_j-1).
\]
By the separable-convex allocation property~\cite{geiger2016greedy},
this condition implies that $\bw$ is a global minimizer of $f$
under the fixed-sum constraint.
Hence $D(\bp\|\bq)\le D(\bp\|\bq')$
for every $\bq'\in\mathcal C_\delta$,
so $\bp\in V_{\mathrm{KL}}(\bq)$.
\end{proof}

\begin{remark}[KLQ computation]
\label{rem:klq-local-search}
Geiger and B\"ocherer give an exact greedy algorithm that allocates counts one at a time for this separable convex problem~\cite{geiger2016greedy}.
Lemma~\ref{lem:klq-cell-local} gives an alternative local-search view, where counts are moved between coordinates by lattice-adjacent swaps until no improving swap remains. The resulting codeword is KL-optimal.
\end{remark}

Figure~\ref{fig:cells-klq} shows the two KLQ cell partitions at
$m=3$, $k=2$ ($K=9$).
The codebook is the same as ELQ at the corresponding $\alpha$,
but the cells differ because KL-bisectors differ from $\ell_2$-bisectors.

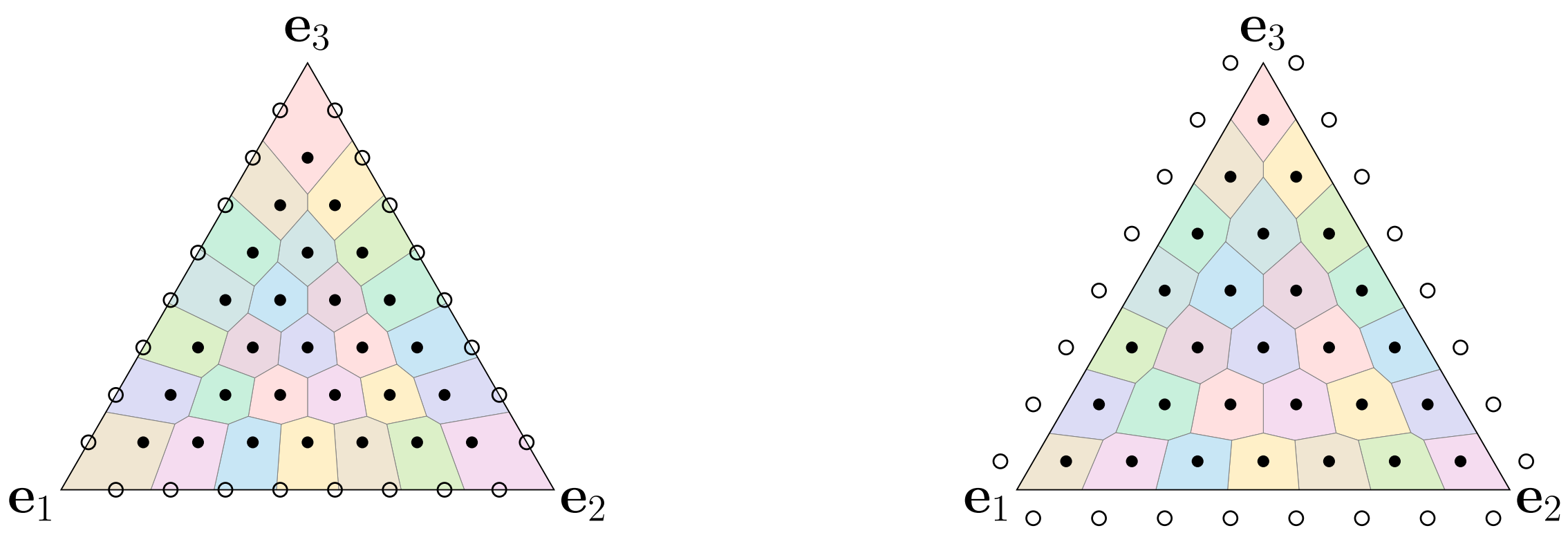
\begin{figure}[!t]
\centering
\begin{subfigure}[!t]{0.4\linewidth}
\centering
\resizebox{\linewidth}{!}{\begin{tikzpicture}[scale=5.5,every node/.append style={font=\Large}]
  \fill[color={rgb,255:red,255;green,224;blue,224}, draw=black!50, line width=0.15pt] (0.59097,0.70847) -- (0.50000,0.86603) -- (0.40903,0.70847) -- (0.50000,0.59941) -- cycle;
  \fill[color={rgb,255:red,255;green,240;blue,200}, draw=black!50, line width=0.15pt] (0.65509,0.59740) -- (0.59097,0.70847) -- (0.50000,0.59941) -- (0.50000,0.56749) -- (0.55448,0.50565) -- cycle;
  \fill[color={rgb,255:red,220;green,240;blue,200}, draw=black!50, line width=0.15pt] (0.55448,0.50565) -- (0.56100,0.46255) -- (0.60816,0.41287) -- (0.71841,0.48772) -- (0.65509,0.59740) -- cycle;
  \fill[color={rgb,255:red,200;green,240;blue,220}, draw=black!50, line width=0.15pt] (0.60816,0.41287) -- (0.62111,0.35860) -- (0.66163,0.32024) -- (0.78159,0.37831) -- (0.71841,0.48772) -- cycle;
  \fill[color={rgb,255:red,200;green,230;blue,245}, draw=black!50, line width=0.15pt] (0.84491,0.26863) -- (0.78159,0.37831) -- (0.66163,0.32024) -- (0.68108,0.25456) -- (0.71515,0.22737) -- cycle;
  \fill[color={rgb,255:red,220;green,220;blue,245}, draw=black!50, line width=0.15pt] (0.90903,0.15756) -- (0.84491,0.26863) -- (0.71515,0.22737) -- (0.74146,0.14927) -- (0.76911,0.13331) -- cycle;
  \fill[color={rgb,255:red,245;green,220;blue,240}, draw=black!50, line width=0.15pt] (1.00000,0.00000) -- (0.90903,0.15756) -- (0.76911,0.13331) -- (0.81807,0.00000) -- cycle;
  \fill[color={rgb,255:red,240;green,230;blue,210}, draw=black!50, line width=0.15pt] (0.50000,0.59941) -- (0.40903,0.70847) -- (0.34491,0.59740) -- (0.44552,0.50565) -- (0.50000,0.56749) -- cycle;
  \fill[color={rgb,255:red,210;green,230;blue,230}, draw=black!50, line width=0.15pt] (0.56100,0.46255) -- (0.55448,0.50565) -- (0.50000,0.56749) -- (0.44552,0.50565) -- (0.43900,0.46255) -- (0.50000,0.41226) -- cycle;
  \fill[color={rgb,255:red,235;green,215;blue,225}, draw=black!50, line width=0.15pt] (0.62111,0.35860) -- (0.60816,0.41287) -- (0.56100,0.46255) -- (0.50000,0.41226) -- (0.50000,0.35801) -- (0.55361,0.31962) -- cycle;
  \fill[color={rgb,255:red,255;green,224;blue,224}, draw=black!50, line width=0.15pt] (0.68108,0.25456) -- (0.66163,0.32024) -- (0.62111,0.35860) -- (0.55361,0.31962) -- (0.56004,0.25401) -- (0.60703,0.22688) -- cycle;
  \fill[color={rgb,255:red,255;green,240;blue,200}, draw=black!50, line width=0.15pt] (0.74146,0.14927) -- (0.71515,0.22737) -- (0.68108,0.25456) -- (0.60703,0.22688) -- (0.62008,0.14891) -- (0.66066,0.13300) -- cycle;
  \fill[color={rgb,255:red,220;green,240;blue,200}, draw=black!50, line width=0.15pt] (0.81807,0.00000) -- (0.76911,0.13331) -- (0.74146,0.14927) -- (0.66066,0.13300) -- (0.68982,0.00000) -- cycle;
  \fill[color={rgb,255:red,200;green,240;blue,220}, draw=black!50, line width=0.15pt] (0.43900,0.46255) -- (0.44552,0.50565) -- (0.34491,0.59740) -- (0.28159,0.48772) -- (0.39184,0.41287) -- cycle;
  \fill[color={rgb,255:red,200;green,230;blue,245}, draw=black!50, line width=0.15pt] (0.50000,0.35801) -- (0.50000,0.41226) -- (0.43900,0.46255) -- (0.39184,0.41287) -- (0.37889,0.35860) -- (0.44639,0.31962) -- cycle;
  \fill[color={rgb,255:red,220;green,220;blue,245}, draw=black!50, line width=0.15pt] (0.56004,0.25401) -- (0.55361,0.31962) -- (0.50000,0.35801) -- (0.44639,0.31962) -- (0.43996,0.25401) -- (0.50000,0.22678) -- cycle;
  \fill[color={rgb,255:red,245;green,220;blue,240}, draw=black!50, line width=0.15pt] (0.62008,0.14891) -- (0.60703,0.22688) -- (0.56004,0.25401) -- (0.50000,0.22678) -- (0.50000,0.14883) -- (0.55348,0.13291) -- cycle;
  \fill[color={rgb,255:red,240;green,230;blue,210}, draw=black!50, line width=0.15pt] (0.68982,-0.00000) -- (0.66066,0.13300) -- (0.62008,0.14891) -- (0.55348,0.13291) -- (0.56317,-0.00000) -- cycle;
  \fill[color={rgb,255:red,210;green,230;blue,230}, draw=black!50, line width=0.15pt] (0.39184,0.41287) -- (0.28159,0.48772) -- (0.21841,0.37831) -- (0.33837,0.32024) -- (0.37889,0.35860) -- cycle;
  \fill[color={rgb,255:red,235;green,215;blue,225}, draw=black!50, line width=0.15pt] (0.43996,0.25401) -- (0.44639,0.31962) -- (0.37889,0.35860) -- (0.33837,0.32024) -- (0.31892,0.25456) -- (0.39297,0.22688) -- cycle;
  \fill[color={rgb,255:red,255;green,224;blue,224}, draw=black!50, line width=0.15pt] (0.50000,0.14883) -- (0.50000,0.22678) -- (0.43996,0.25401) -- (0.39297,0.22688) -- (0.37992,0.14891) -- (0.44652,0.13291) -- cycle;
  \fill[color={rgb,255:red,255;green,240;blue,200}, draw=black!50, line width=0.15pt] (0.56317,0.00000) -- (0.55348,0.13291) -- (0.50000,0.14883) -- (0.44652,0.13291) -- (0.43683,0.00000) -- cycle;
  \fill[color={rgb,255:red,220;green,240;blue,200}, draw=black!50, line width=0.15pt] (0.33837,0.32024) -- (0.21841,0.37831) -- (0.15509,0.26863) -- (0.28485,0.22737) -- (0.31892,0.25456) -- cycle;
  \fill[color={rgb,255:red,200;green,240;blue,220}, draw=black!50, line width=0.15pt] (0.37992,0.14891) -- (0.39297,0.22688) -- (0.31892,0.25456) -- (0.28485,0.22737) -- (0.25854,0.14927) -- (0.33934,0.13300) -- cycle;
  \fill[color={rgb,255:red,200;green,230;blue,245}, draw=black!50, line width=0.15pt] (0.43683,0.00000) -- (0.44652,0.13291) -- (0.37992,0.14891) -- (0.33934,0.13300) -- (0.31018,-0.00000) -- cycle;
  \fill[color={rgb,255:red,220;green,220;blue,245}, draw=black!50, line width=0.15pt] (0.28485,0.22737) -- (0.15509,0.26863) -- (0.09097,0.15756) -- (0.23089,0.13331) -- (0.25854,0.14927) -- cycle;
  \fill[color={rgb,255:red,245;green,220;blue,240}, draw=black!50, line width=0.15pt] (0.31018,0.00000) -- (0.33934,0.13300) -- (0.25854,0.14927) -- (0.23089,0.13331) -- (0.18193,0.00000) -- cycle;
  \fill[color={rgb,255:red,240;green,230;blue,210}, draw=black!50, line width=0.15pt] (0.23089,0.13331) -- (0.09097,0.15756) -- (0.00000,0.00000) -- (0.18193,0.00000) -- cycle;
  \draw[black, line width=0.4pt] (0.00000,0.00000) -- (1.00000,0.00000) -- (0.50000,0.86603) -- cycle;
  \draw[black, line width=0.6pt] (0.55556,0.76980) circle (0.014);
  \draw[black, line width=0.6pt] (0.61111,0.67358) circle (0.014);
  \draw[black, line width=0.6pt] (0.66667,0.57735) circle (0.014);
  \draw[black, line width=0.6pt] (0.72222,0.48113) circle (0.014);
  \draw[black, line width=0.6pt] (0.77778,0.38490) circle (0.014);
  \draw[black, line width=0.6pt] (0.83333,0.28868) circle (0.014);
  \draw[black, line width=0.6pt] (0.88889,0.19245) circle (0.014);
  \draw[black, line width=0.6pt] (0.94444,0.09623) circle (0.014);
  \draw[black, line width=0.6pt] (0.44444,0.76980) circle (0.014);
  \draw[black, line width=0.6pt] (0.88889,0.00000) circle (0.014);
  \draw[black, line width=0.6pt] (0.38889,0.67358) circle (0.014);
  \draw[black, line width=0.6pt] (0.77778,0.00000) circle (0.014);
  \draw[black, line width=0.6pt] (0.33333,0.57735) circle (0.014);
  \draw[black, line width=0.6pt] (0.66667,0.00000) circle (0.014);
  \draw[black, line width=0.6pt] (0.27778,0.48113) circle (0.014);
  \draw[black, line width=0.6pt] (0.55556,0.00000) circle (0.014);
  \draw[black, line width=0.6pt] (0.22222,0.38490) circle (0.014);
  \draw[black, line width=0.6pt] (0.44444,0.00000) circle (0.014);
  \draw[black, line width=0.6pt] (0.16667,0.28868) circle (0.014);
  \draw[black, line width=0.6pt] (0.33333,0.00000) circle (0.014);
  \draw[black, line width=0.6pt] (0.11111,0.19245) circle (0.014);
  \draw[black, line width=0.6pt] (0.22222,0.00000) circle (0.014);
  \draw[black, line width=0.6pt] (0.05556,0.09623) circle (0.014);
  \draw[black, line width=0.6pt] (0.11111,0.00000) circle (0.014);
  \fill[black] (0.50000,0.67358) circle (0.012);
  \fill[black] (0.55556,0.57735) circle (0.012);
  \fill[black] (0.61111,0.48113) circle (0.012);
  \fill[black] (0.66667,0.38490) circle (0.012);
  \fill[black] (0.72222,0.28868) circle (0.012);
  \fill[black] (0.77778,0.19245) circle (0.012);
  \fill[black] (0.83333,0.09623) circle (0.012);
  \fill[black] (0.44444,0.57735) circle (0.012);
  \fill[black] (0.50000,0.48113) circle (0.012);
  \fill[black] (0.55556,0.38490) circle (0.012);
  \fill[black] (0.61111,0.28868) circle (0.012);
  \fill[black] (0.66667,0.19245) circle (0.012);
  \fill[black] (0.72222,0.09623) circle (0.012);
  \fill[black] (0.38889,0.48113) circle (0.012);
  \fill[black] (0.44444,0.38490) circle (0.012);
  \fill[black] (0.50000,0.28868) circle (0.012);
  \fill[black] (0.55556,0.19245) circle (0.012);
  \fill[black] (0.61111,0.09623) circle (0.012);
  \fill[black] (0.33333,0.38490) circle (0.012);
  \fill[black] (0.38889,0.28868) circle (0.012);
  \fill[black] (0.44444,0.19245) circle (0.012);
  \fill[black] (0.50000,0.09623) circle (0.012);
  \fill[black] (0.27778,0.28868) circle (0.012);
  \fill[black] (0.33333,0.19245) circle (0.012);
  \fill[black] (0.38889,0.09623) circle (0.012);
  \fill[black] (0.22222,0.19245) circle (0.012);
  \fill[black] (0.27778,0.09623) circle (0.012);
  \fill[black] (0.16667,0.09623) circle (0.012);
  \node[anchor=center] at (-0.06000,-0.03000) {$\mathbf{e}_1$};
  \node[anchor=center] at (1.06000,-0.03000) {$\mathbf{e}_2$};
  \node[anchor=center] at (0.50000,0.92603) {$\mathbf{e}_3$};
\end{tikzpicture}}
\caption{$\alpha=\delta=1/K$}
\label{fig:cells-klq-sym}
\end{subfigure}\hfill
\begin{subfigure}[!t]{0.4\linewidth}
\centering
\resizebox{\linewidth}{!}{\begin{tikzpicture}[scale=5.5,every node/.append style={font=\Large}]
  \fill[color={rgb,255:red,255;green,224;blue,224}, draw=black!50, line width=0.15pt] (0.50000,0.86603) -- (0.43401,0.75172) -- (0.50000,0.66407) -- (0.56599,0.75172) -- cycle;
  \fill[color={rgb,255:red,255;green,240;blue,200}, draw=black!50, line width=0.15pt] (0.64102,0.62177) -- (0.56599,0.75172) -- (0.50000,0.66407) -- (0.50000,0.63430) -- (0.56670,0.54968) -- cycle;
  \fill[color={rgb,255:red,220;green,240;blue,200}, draw=black!50, line width=0.15pt] (0.56670,0.54968) -- (0.57649,0.50329) -- (0.63113,0.43835) -- (0.71378,0.49575) -- (0.64102,0.62177) -- cycle;
  \fill[color={rgb,255:red,200;green,240;blue,220}, draw=black!50, line width=0.15pt] (0.78622,0.37028) -- (0.71378,0.49575) -- (0.63113,0.43835) -- (0.65040,0.37551) -- (0.69519,0.32740) -- cycle;
  \fill[color={rgb,255:red,200;green,230;blue,245}, draw=black!50, line width=0.15pt] (0.85898,0.24425) -- (0.78622,0.37028) -- (0.69519,0.32740) -- (0.72410,0.24761) -- (0.75939,0.21594) -- cycle;
  \fill[color={rgb,255:red,220;green,220;blue,245}, draw=black!50, line width=0.15pt] (0.93401,0.11431) -- (0.85898,0.24425) -- (0.75939,0.21594) -- (0.79932,0.11586) -- (0.82510,0.10098) -- cycle;
  \fill[color={rgb,255:red,245;green,220;blue,240}, draw=black!50, line width=0.15pt] (1.00000,0.00000) -- (0.93401,0.11431) -- (0.82510,0.10098) -- (0.86801,0.00000) -- cycle;
  \fill[color={rgb,255:red,240;green,230;blue,210}, draw=black!50, line width=0.15pt] (0.50000,0.66407) -- (0.43401,0.75172) -- (0.35898,0.62177) -- (0.43330,0.54968) -- (0.50000,0.63430) -- cycle;
  \fill[color={rgb,255:red,210;green,230;blue,230}, draw=black!50, line width=0.15pt] (0.57649,0.50329) -- (0.56670,0.54968) -- (0.50000,0.63430) -- (0.43330,0.54968) -- (0.42351,0.50329) -- (0.50000,0.43651) -- cycle;
  \fill[color={rgb,255:red,235;green,215;blue,225}, draw=black!50, line width=0.15pt] (0.65040,0.37551) -- (0.63113,0.43835) -- (0.57649,0.50329) -- (0.50000,0.43651) -- (0.50000,0.37371) -- (0.56419,0.32573) -- cycle;
  \fill[color={rgb,255:red,255;green,224;blue,224}, draw=black!50, line width=0.15pt] (0.72410,0.24761) -- (0.69519,0.32740) -- (0.65040,0.37551) -- (0.56419,0.32573) -- (0.57364,0.24616) -- (0.62803,0.21476) -- cycle;
  \fill[color={rgb,255:red,255;green,240;blue,200}, draw=black!50, line width=0.15pt] (0.79932,0.11586) -- (0.75939,0.21594) -- (0.72410,0.24761) -- (0.62803,0.21476) -- (0.64762,0.11513) -- (0.69269,0.10040) -- cycle;
  \fill[color={rgb,255:red,220;green,240;blue,200}, draw=black!50, line width=0.15pt] (0.86801,0.00000) -- (0.82510,0.10098) -- (0.79932,0.11586) -- (0.69269,0.10040) -- (0.71796,-0.00000) -- cycle;
  \fill[color={rgb,255:red,200;green,240;blue,220}, draw=black!50, line width=0.15pt] (0.43330,0.54968) -- (0.35898,0.62177) -- (0.28622,0.49575) -- (0.36887,0.43835) -- (0.42351,0.50329) -- cycle;
  \fill[color={rgb,255:red,200;green,230;blue,245}, draw=black!50, line width=0.15pt] (0.50000,0.37371) -- (0.50000,0.43651) -- (0.42351,0.50329) -- (0.36887,0.43835) -- (0.34960,0.37551) -- (0.43581,0.32573) -- cycle;
  \fill[color={rgb,255:red,220;green,220;blue,245}, draw=black!50, line width=0.15pt] (0.57364,0.24616) -- (0.56419,0.32573) -- (0.50000,0.37371) -- (0.43581,0.32573) -- (0.42636,0.24616) -- (0.50000,0.21456) -- cycle;
  \fill[color={rgb,255:red,245;green,220;blue,240}, draw=black!50, line width=0.15pt] (0.64762,0.11513) -- (0.62803,0.21476) -- (0.57364,0.24616) -- (0.50000,0.21456) -- (0.50000,0.11501) -- (0.56405,0.10027) -- cycle;
  \fill[color={rgb,255:red,240;green,230;blue,210}, draw=black!50, line width=0.15pt] (0.71796,-0.00000) -- (0.69269,0.10040) -- (0.64762,0.11513) -- (0.56405,0.10027) -- (0.57244,-0.00000) -- cycle;
  \fill[color={rgb,255:red,210;green,230;blue,230}, draw=black!50, line width=0.15pt] (0.36887,0.43835) -- (0.28622,0.49575) -- (0.21378,0.37028) -- (0.30481,0.32740) -- (0.34960,0.37551) -- cycle;
  \fill[color={rgb,255:red,235;green,215;blue,225}, draw=black!50, line width=0.15pt] (0.42636,0.24616) -- (0.43581,0.32573) -- (0.34960,0.37551) -- (0.30481,0.32740) -- (0.27590,0.24761) -- (0.37197,0.21476) -- cycle;
  \fill[color={rgb,255:red,255;green,224;blue,224}, draw=black!50, line width=0.15pt] (0.50000,0.11501) -- (0.50000,0.21456) -- (0.42636,0.24616) -- (0.37197,0.21476) -- (0.35238,0.11513) -- (0.43595,0.10027) -- cycle;
  \fill[color={rgb,255:red,255;green,240;blue,200}, draw=black!50, line width=0.15pt] (0.57244,0.00000) -- (0.56405,0.10027) -- (0.50000,0.11501) -- (0.43595,0.10027) -- (0.42756,0.00000) -- cycle;
  \fill[color={rgb,255:red,220;green,240;blue,200}, draw=black!50, line width=0.15pt] (0.30481,0.32740) -- (0.21378,0.37028) -- (0.14102,0.24425) -- (0.24061,0.21594) -- (0.27590,0.24761) -- cycle;
  \fill[color={rgb,255:red,200;green,240;blue,220}, draw=black!50, line width=0.15pt] (0.35238,0.11513) -- (0.37197,0.21476) -- (0.27590,0.24761) -- (0.24061,0.21594) -- (0.20068,0.11586) -- (0.30731,0.10040) -- cycle;
  \fill[color={rgb,255:red,200;green,230;blue,245}, draw=black!50, line width=0.15pt] (0.42756,0.00000) -- (0.43595,0.10027) -- (0.35238,0.11513) -- (0.30731,0.10040) -- (0.28204,-0.00000) -- cycle;
  \fill[color={rgb,255:red,220;green,220;blue,245}, draw=black!50, line width=0.15pt] (0.24061,0.21594) -- (0.14102,0.24425) -- (0.06599,0.11431) -- (0.17490,0.10098) -- (0.20068,0.11586) -- cycle;
  \fill[color={rgb,255:red,245;green,220;blue,240}, draw=black!50, line width=0.15pt] (0.28204,0.00000) -- (0.30731,0.10040) -- (0.20068,0.11586) -- (0.17490,0.10098) -- (0.13199,0.00000) -- cycle;
  \fill[color={rgb,255:red,240;green,230;blue,210}, draw=black!50, line width=0.15pt] (0.17490,0.10098) -- (0.06599,0.11431) -- (0.00000,0.00000) -- (0.13199,0.00000) -- cycle;
  \draw[black, line width=0.4pt] (0.00000,0.00000) -- (1.00000,0.00000) -- (0.50000,0.86603) -- cycle;
  \draw[black, line width=0.6pt] (0.56667,0.86603) circle (0.014);
  \draw[black, line width=0.6pt] (0.63333,0.75056) circle (0.014);
  \draw[black, line width=0.6pt] (0.70000,0.63509) circle (0.014);
  \draw[black, line width=0.6pt] (0.76667,0.51962) circle (0.014);
  \draw[black, line width=0.6pt] (0.83333,0.40415) circle (0.014);
  \draw[black, line width=0.6pt] (0.90000,0.28868) circle (0.014);
  \draw[black, line width=0.6pt] (0.96667,0.17321) circle (0.014);
  \draw[black, line width=0.6pt] (1.03333,0.05774) circle (0.014);
  \draw[black, line width=0.6pt] (0.43333,0.86603) circle (0.014);
  \draw[black, line width=0.6pt] (0.96667,-0.05774) circle (0.014);
  \draw[black, line width=0.6pt] (0.36667,0.75056) circle (0.014);
  \draw[black, line width=0.6pt] (0.83333,-0.05774) circle (0.014);
  \draw[black, line width=0.6pt] (0.30000,0.63509) circle (0.014);
  \draw[black, line width=0.6pt] (0.70000,-0.05774) circle (0.014);
  \draw[black, line width=0.6pt] (0.23333,0.51962) circle (0.014);
  \draw[black, line width=0.6pt] (0.56667,-0.05774) circle (0.014);
  \draw[black, line width=0.6pt] (0.16667,0.40415) circle (0.014);
  \draw[black, line width=0.6pt] (0.43333,-0.05774) circle (0.014);
  \draw[black, line width=0.6pt] (0.10000,0.28868) circle (0.014);
  \draw[black, line width=0.6pt] (0.30000,-0.05774) circle (0.014);
  \draw[black, line width=0.6pt] (0.03333,0.17321) circle (0.014);
  \draw[black, line width=0.6pt] (0.16667,-0.05774) circle (0.014);
  \draw[black, line width=0.6pt] (-0.03333,0.05774) circle (0.014);
  \draw[black, line width=0.6pt] (0.03333,-0.05774) circle (0.014);
  \fill[black] (0.50000,0.75056) circle (0.012);
  \fill[black] (0.56667,0.63509) circle (0.012);
  \fill[black] (0.63333,0.51962) circle (0.012);
  \fill[black] (0.70000,0.40415) circle (0.012);
  \fill[black] (0.76667,0.28868) circle (0.012);
  \fill[black] (0.83333,0.17321) circle (0.012);
  \fill[black] (0.90000,0.05774) circle (0.012);
  \fill[black] (0.43333,0.63509) circle (0.012);
  \fill[black] (0.50000,0.51962) circle (0.012);
  \fill[black] (0.56667,0.40415) circle (0.012);
  \fill[black] (0.63333,0.28868) circle (0.012);
  \fill[black] (0.70000,0.17321) circle (0.012);
  \fill[black] (0.76667,0.05774) circle (0.012);
  \fill[black] (0.36667,0.51962) circle (0.012);
  \fill[black] (0.43333,0.40415) circle (0.012);
  \fill[black] (0.50000,0.28868) circle (0.012);
  \fill[black] (0.56667,0.17321) circle (0.012);
  \fill[black] (0.63333,0.05774) circle (0.012);
  \fill[black] (0.30000,0.40415) circle (0.012);
  \fill[black] (0.36667,0.28868) circle (0.012);
  \fill[black] (0.43333,0.17321) circle (0.012);
  \fill[black] (0.50000,0.05774) circle (0.012);
  \fill[black] (0.23333,0.28868) circle (0.012);
  \fill[black] (0.30000,0.17321) circle (0.012);
  \fill[black] (0.36667,0.05774) circle (0.012);
  \fill[black] (0.16667,0.17321) circle (0.012);
  \fill[black] (0.23333,0.05774) circle (0.012);
  \fill[black] (0.10000,0.05774) circle (0.012);
  \node[anchor=center] at (-0.06000,-0.03000) {$\mathbf{e}_1$};
  \node[anchor=center] at (1.06000,-0.03000) {$\mathbf{e}_2$};
  \node[anchor=center] at (0.50000,0.92603) {$\mathbf{e}_3$};
\end{tikzpicture}}
\caption{$\alpha=\alpha^\star$}
\label{fig:cells-klq-hex}
\end{subfigure}
\caption{KLQ (KL-Voronoi) cell partitions at $m=3$, $k=2$
($K=9$, 28 codewords). Cells are convex polytopes
(Lemma~\ref{lem:kl-convex}).}
\label{fig:cells-klq}
\end{figure}
\end{vlong}

\subsection{Inradius and stability}
\label{sec:klq-smooth}

By Lemma~\ref{lem:klq-cell-local}, only lattice-adjacent KL bisectors
need to be considered when finding the inradius.

\begin{definition}
\label{def:klq-inradius}
For $\bq\in\mathcal C_\delta$, define
$\rho(\bq)=\mathrm{dist}\bigl(\bq,\partial V_{\mathrm{KL}}(\bq)\bigr)$,
where the distance and boundary are taken in the affine hull of
$\Delta_{m-1}$.
\end{definition}

The nearest boundary of a KLQ cell can come from either
a lattice-adjacent KL bisector or a simplex facet.

\begin{proposition}[KLQ inradius at $\alpha=\delta=1/K$]
\label{prop:klq-inradius}
At $\alpha=\delta=1/K$, with $K=m(k+1)$, the KLQ inradius satisfies
\begin{equation}
\rho(m,K)\;\triangleq\;\min\nolimits_{\bq\in\mathcal{C}_\delta}\rho(\bq)
\;\ge\;
\frac{1-\ln 2}{\ln 2}
\sqrt{\tfrac{m}{m-1}}\cdot\frac{1}{K},
\end{equation}
and the bound is asymptotically tight as $k\to\infty$
(equivalently, $K\to\infty$ along $K=m(k+1)$).
\end{proposition}

\begin{vlong}
\begin{proof}[Proof sketch]
At $\alpha=\delta=1/K$, write
$u_i\triangleq Kq_i=w_i+1\ge1$.
By Lemmas~\ref{lem:kl-convex} and~\ref{lem:klq-cell-local},
$\rho(\bq)$ is the smaller of the distance to a lattice-adjacent
KL bisector and the distance to a simplex facet.
The calculation in Appendix~\ref{app:inradius} shows that every
adjacent-bisector distance is at least
\[
\frac{1-\ln 2}{\ln 2}
\sqrt{\frac{m}{m-1}}\frac{1}{K},
\]
and corner codewords approach this bound as $k\to\infty$.
The simplex-facet distance is at least
$(1/K)\sqrt{m/(m-1)}$, which is larger because
$(1-\ln 2)/\ln 2<1$.
\end{proof}
\end{vlong}

\begin{theorem}[KLQ quantization stability]
\label{thm:klq-smooth}
At $\alpha=\delta=1/K$, with $K=m(k+1)$,
for all $\varepsilon\in[0,\rho/2]$,
\[
S(\varepsilon)\ge S_{\mathrm{margin}}(\varepsilon)
\ge(1-2\varepsilon/\rho)^{m-1},
\qquad
\rho=\min\nolimits_{\bq\in\mathcal{C}_\delta}\rho(\bq).
\]
\end{theorem}

\begin{vlong}
\begin{proof}
The proof uses the same scaling argument as
Theorem~\ref{thm:elq-smooth}.
KLQ cells are convex by Lemma~\ref{lem:kl-convex},
and Proposition~\ref{prop:klq-inradius} gives the required inradius bound.
\end{proof}
\end{vlong}

\subsection{Worst-case KL distortion}
\label{sec:klq-kl-bound}

\begin{vlong}
In this subsection, we take $\alpha=\delta=1/K$ with $K=m(k+1)$, so that $\mathcal C_\delta=\Sint_K^m$. For each $\bp\in\Delta_{m-1}$, we construct a codeword $\bq_0\in\mathcal C_\delta$ with a bound on $\max_{i:p_i>0}p_i/q_{0,i}$. KL-optimality then gives the corresponding bound for $\phi_{\mathrm{KL}}(\bp)$.

\begin{theorem}
\label{thm:prop-bound}
For every boundary type $\hat\bp\in\hat{\mathbb S}_K^m$,
\begin{equation}
R^*(\hat\bp)\triangleq\min_{\bq\in\Sint_K^m}\max_{i:\hat p_i>0}\frac{\hat p_i}{q_i}\;\le\;\frac{K}{K-m+1},
\end{equation}
with equality at $\hat\bp=\be_1$, attained by $\bq^*=\tfrac{1}{K}(K-m+1,1,\ldots,1)$.
\end{theorem}

\begin{proof}
Set $R=K/(K-m+1)$ and write any type $\bq\in\Sint_K^m$ via its count vector $\bu\triangleq K\bq\in\mathbb{Z}_{\ge 1}^m$ with $\sum_i u_i=K$; similarly $\hat\bv\triangleq K\hat\bp$. Define
\[
\tilde u_i=\begin{cases}1 & \hat v_i=0,\\ \ceil{\hat v_i(K-m+1)/K} & \hat v_i\ge 1.\end{cases}
\]
Then $\tilde u_i\ge 1$. For $\hat v_i\ge 1$, $\tilde u_i\ge\hat v_i(K-m+1)/K$, so $\hat v_i/\tilde u_i\le R$. Let $s=|\{i:\hat v_i\ge1\}|$. Then
\begin{equation*}
\begin{split}
\sum_i\tilde u_i
&<(m-s)+\sum_{\hat v_i\ge1}
\bigl(\hat v_i(K-m+1)/K+1\bigr)\\
&=(m-s)+(K-m+1)+s=K+1.
\end{split}
\end{equation*}
Since $\sum_i\tilde u_i$ is an integer, $\sum_i\tilde u_i\le K$. Distribute the remaining $K-\sum_i\tilde u_i$ counts among the coordinates; this only decreases ratios. Set $\tilde\bq=\tilde\bu/K\in\Sint_K^m$. For $\hat\bp=\be_1$, every $\bq\in\Sint_K^m$ has $Kq_1\le K-(m-1)$, forcing $\hat p_1/q_1\ge R$.
\end{proof}

The same construction extends from boundary types to all $\bp\in\Delta_{m-1}$.

\begin{corollary}
\label{cor:prop-continuous}
For every $\bp\in\Delta_{m-1}$, there exists $\bq\in\mathcal{C}_\delta$ with $\max_{i:p_i>0}\frac{p_i}{q_i}\le \frac{K}{K-m+1}$.
\end{corollary}

\begin{proof}
Set $u_i=\max(1,\ceil{p_i(K-m+1)})$. As in the proof of Theorem~\ref{thm:prop-bound}, $\sum_i u_i\le K$ and $p_iK/u_i\le K/(K-m+1)$ for every $i$ with $p_i>0$. Distribute the remaining $K-\sum_i u_i$ counts among the coordinates and set $\bq=\bu/K$. Increasing the counts can only decrease the ratios, so the same bound holds.
\end{proof}

Using Corollary~\ref{cor:prop-continuous}, Equation~\eqref{eq:kl-from-ratio}, and KL-optimality gives the following worst-case bound.

\begin{theorem}
\label{thm:klq-kl}
At $\alpha=\delta=1/K$, with $K=m(k+1)$,
\[
\max_{\bp\in\Delta_{m-1}}
D(\bp\|\phi_{\mathrm{KL}}(\bp))
=
\ln\frac{K}{K-m+1}.
\]
\end{theorem}

\begin{proof}
By Corollary~\ref{cor:prop-continuous} there exists $\bq_0\in\mathcal{C}_\delta$ with $\max_{i:p_i>0}\frac{p_i}{q_{0,i}}\le \frac{K}{K-m+1}$. By \eqref{eq:kl-from-ratio}, $D(\bp\|\bq_0)\le\ln\bigl(\max_{i:p_i>0}\frac{p_i}{q_{0,i}}\bigr)\le\ln\frac{K}{K-m+1}$. By the KL-optimality of $\phi_{\mathrm{KL}}$,
\[
D(\bp\|\phi_{\mathrm{KL}}(\bp))\le D(\bp\|\bq_0)\le\ln\frac{K}{K-m+1}.
\]
Equality is attained at $\bp=\be_1$, since every $\bq\in\mathcal C_\delta$ satisfies $q_1\le(K-m+1)/K$, and this value is attained by $\bq^*=(K-m+1,1,\ldots,1)/K$.
\end{proof}

For distributions with $p_i>0$ for all $i$, this is the same $K$-type KL approximation problem studied by Geiger and B\"ocherer \cite{geiger2016greedy}. Their distribution-dependent bound can be tighter away from the simplex boundary. For example, when $p_i\ge 1/K$ for all $i$, they obtain $D(\bp\|\phi_{\rm KL}(\bp))\le m\ln 2/(2K)$. As $\bp$ remains in the simplex interior and approaches a vertex, their bound approaches $\ln\frac{K}{K-m+1}+\frac{\ln 2}{2(K-m+1)}$, while our distortion approaches $\ln\frac{K}{K-m+1}$, which is the exact worst-case distortion over the full simplex. Their proof uses a reverse I-projection and a variational-distance bound, while our exact result follows from a direct construction of a positive $K$-type and a matching lower bound at a simplex vertex.

Adler, Tang, and Polyanskiy studied a different quantizer based on entrywise companding and proved a worst-case KL bound of order $N_{\mathrm{A}}^{-2}\log^2 m$, where $N_{\mathrm{A}}$ is the number of quantization levels per coordinate~\cite{adler2023efficient}. Their representation uses $m\lceil\log_2N_{\mathrm{A}}\rceil$ bits, while an index of the symmetric lattice codebook here requires $\left\lceil\log_2\binom{K-1}{m-1}\right\rceil$ bits, so $N_{\mathrm{A}}$ and $K$ should not be compared directly. Our use of the lattice is motivated by the quantization stability analysis in Section~\ref{sec:klq-smooth}, rather than KL distortion alone.

For $\bp$ lying in a lattice Voronoi cell unclipped by the simplex boundary, KL-optimality of $\phi_{\mathrm{KL}}$ together with Lemma~\ref{lem:elq-lattice-cell} gives the tighter bound $D(\bp\|\phi_{\mathrm{KL}}(\bp))\le 1/[4(k+1)]$ (at $\alpha=\delta=1/K$).
\end{vlong}

\begin{vlong}
\subsection{Lloyd algorithm and centroidal optimality}
\label{sec:klq-lloyd}

We also apply Lloyd's algorithm~\cite{lloyd1982least} to reduce the average KL distortion by alternating between a cell update and a centroid update.

Let $\bp\sim\mathrm{Unif}(\Delta_{m-1})$ and write $J(\mathcal C) = \mathbb{E}\!\left[ \min_{\bq\in\mathcal C}D(\bp\|\bq) \right]$.

\paragraph{Lloyd iteration}
Given a codebook $\mathcal{C}^{(t)}=\{\bq_1^{(t)},\ldots,\bq_N^{(t)}\}$:
\begin{itemize}
  \item \emph{Cell update}: form $V_i^{(t)}=\{\bp\in\Delta_{m-1}:D(\bp\|\bq_i^{(t)})\le D(\bp\|\bq_j^{(t)})\ \forall j\}$.
  \item \emph{Centroid update}: set $\bq_i^{(t+1)}=\arg\min_{\bq\in\Delta^\circ_{m-1}}\mathbb{E}[D(\bp\|\bq)\mid\bp\in V_i^{(t)}]$.
\end{itemize}
For KL divergence, the minimizing codeword is the conditional mean of $\bp$ in the cell, a standard result for Bregman divergences~\cite{JMLR:v6:banerjee05b}, so $\bq_i^{(t+1)}=\mathbb{E}[\bp\mid\bp\in V_i^{(t)}]$.

Under the uniform distribution on $\Delta_{m-1}$, this conditional mean is the geometric centroid of the cell.

\paragraph{Numerical observation}

Each Lloyd sweep is non-increasing in $J(\mathcal C)$, so the sequence of objective values converges. Starting from $\mathcal C^{(0)}=\Sint_K^m$ at $m=3$, $k=3$ ($K=12$), the iteration stops after about $70$ sweeps. The smallest coordinate of the final codebook is $0.0275$, close to the heuristic value $\alpha^\star\approx0.0262$ from Section~\ref{sec:elq-alpha-heuristic}. The Lloyd-optimized codebook is not restricted to the lattice family $\mathcal C_\delta$, so we use this only as a numerical comparison; the analytical KLQ bounds above apply to the lattice codebook. %

\end{vlong}

\section{Simulation}
\label{sec:sim}

\begin{vlong}

We evaluate the stability and KL distortion of the studied quantizers. Source distributions $\boldsymbol{\theta}$ are drawn from $\mathrm{Dir}(1,\ldots,1)$ on alphabet size $m=20$; for each source, we generate an i.i.d.\ sequence of length $L=1000$, compute its empirical type $\bp$, and delete $n_{\mathrm{del}}\in\{1,2,5,10,20,50,100\}$ symbols uniformly at random to obtain $\bp'$.

Given $\boldsymbol{\theta}$, the empirical type $\bp_L$ converges almost surely to $\boldsymbol{\theta}$ as $L\to\infty$ by the strong law of large numbers. To show that the types are uniformly distributed, let $\mathbf{c}=(c_1,\ldots,c_m)$ be a count vector with $c_i\ge0$ and $\sum_i c_i=L$. Integrating the multinomial probability over $\boldsymbol{\theta}\sim\mathrm{Dir}(1,\ldots,1)$ gives
\[
\Pr(\bp=\mathbf{c}/L)=\frac{L!(m-1)!}{(L+m-1)!}=\binom{L+m-1}{m-1}^{-1}.
\]
This probability is the same for every count vector, so $\bp$ is uniform over all length-$L$ types. Let $\mathbf{r}$ be the empirical type of the deleted symbols. Then $\bp=\frac{L-n_{\mathrm{del}}}{L}\bp'+\frac{n_{\mathrm{del}}}{L}\mathbf{r}$, giving
\[
\mathrm{TV}(\bp,\bp')=\frac{n_{\mathrm{del}}}{L}\mathrm{TV}(\mathbf{r},\bp')\le\frac{n_{\mathrm{del}}}{L}.
\]
Unlike the continuous model in (2), random deletion does not sample $\bp'$ uniformly from a TV ball.

We use 500 original sequences and three deletion trials for each $n_{\mathrm{del}}$. We fix $k=1$ (hence $K=40$) and compare ELQ and KLQ at $\alpha=\delta=1/K$. We also compare with the approximate minimax compander (AMC) of Adler, Tang, and Polyanskiy~\cite{adler2023efficient}, using $N_{\mathrm A}=4$ quantization levels per coordinate. AMC uses midpoint decoding followed by normalization~\cite{adler2023efficient}, with zero probabilities assigned to the first bin. We include canonical Huffman coding applied directly to $\bp$ as a baseline.

Under random deletions, we measure codebook stability $S_{\mathrm{cb}}^\phi(n_{\mathrm{del}})=\Pr[\mathrm{CH}(\phi(\bp))=\mathrm{CH}(\phi(\bp'))]$, where $\phi$ is ELQ, KLQ, or AMC. For the canonical Huffman baseline, $\phi$ is the identity map. For ELQ and KLQ, let $\rho_\phi$ be a lower bound on the inradii of their quantization cells. Let $S_{\mathrm{del}}^\phi(n_{\mathrm{del}})=\Pr[\phi(\bp)=\phi(\bp')]$ denote quantization stability under deletions. Theorem~\ref{thm:finite-type-stability} in Appendix~\ref{app:type-stability} gives
\[
S_{\mathrm{cb}}^\phi(n_{\mathrm{del}})\ge S_{\mathrm{del}}^\phi(n_{\mathrm{del}})\ge\left[1-\frac{2(n_{\mathrm{del}}+\sqrt{m})}{L\rho_\phi}\right]_+^{m-1},
\]
where $[x]_+=\max\{x,0\}$.

\IfFileExists{figures/fig2_codebook_stability.pdf}{%
\begin{figure}[tb]
\centering
\begin{minipage}[c]{0.48\linewidth}
    \centering
    \includegraphics[width=\linewidth]{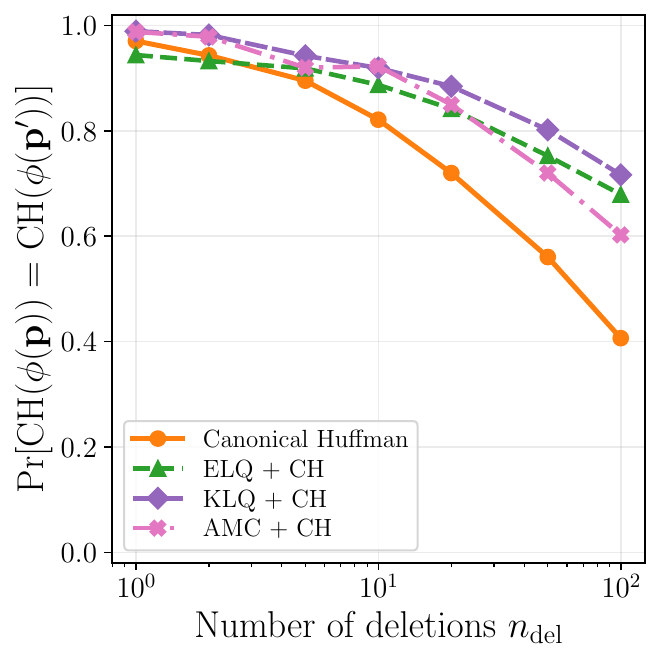}
\end{minipage}\hfill
\begin{minipage}[c]{0.50\linewidth}
    \centering
    \footnotesize
    \setlength{\tabcolsep}{2pt}
    \renewcommand{\arraystretch}{1.15}

    \captionof{table}{KL distortion (nats).}
    \label{tab:kl-distortion}
    \vspace{-0.3em}
    \begin{tabular}{@{}lrrr@{}}
    \toprule
    Method & Mean & Max & Bound \\
    \midrule
    ELQ & 0.0760 & 0.6444 & 3.6889 \\
    KLQ & 0.0732 & 0.6444 & 0.6444 \\
    AMC & 0.0815 & 0.1820 & -- \\
    \bottomrule
    \end{tabular}

    \vspace{1em}

    \captionof{table}{Average bits per sequence.}
    \label{tab:bit-length}
    \vspace{-0.3em}
    \begin{tabular}{@{}lrrr@{}}
    \toprule
    Method & Payload & Header & Total \\
    \midrule
    CH & 3762.6 & 100 & 3862.6 \\
    ELQ+CH & 3916.9 & 37 & 3953.9 \\
    KLQ+CH & 3925.2 & 37 & 3962.2 \\
    AMC+CH & 3888.1 & 40 & 3928.1 \\
    \bottomrule
    \end{tabular}
\end{minipage}

\caption{Codebook stability under random deletions ($m=20$, $L=1000$), comparing canonical Huffman (CH) with ELQ, KLQ, and AMC followed by CH. We use $k=1$ ($K=40$) for ELQ and KLQ.}
\label{fig:codebook-stability}
\end{figure}
}{%
\par\noindent\textit{[Stability figure not yet generated.]}\par\medskip
}

The bound provides a non-asymptotic guarantee for arbitrary deletion
patterns under the uniform-type model. For fixed $m$ and $\phi$,
the bound approaches $1$ as $L \to \infty$ if $n_{\mathrm{del}} = o(L)$. We also measure quantizer distortion $D(\bp\|\phi(\bp))$, which is independent of $n_{\mathrm{del}}$. Table~\ref{tab:kl-distortion} reports the mean KL distortion over the 500 original types, the empirical maximum over all simplex vertices and 5000 additional samples from $\mathrm{Dir}(1,\ldots,1)$, and the theoretical bounds. For ELQ, we use the tighter of the global bound in Lemma~\ref{lem:elq-l2-coarse-gen} and the general bound $D(\bp\|\bq)\le\ln(1/\alpha)$. For KLQ, we use the exact worst-case bound in Theorem~\ref{thm:klq-kl}. For AMC, we report only the empirical mean and empirical worst-case KL distortion because its published nonasymptotic worst-case bound does not apply at $m=20$ and $N_{\mathrm A}=4$.

Table~\ref{tab:bit-length} reports the average payload, header, and total bits for the 500 original sequences. We use 100 bits to store the 20 canonical Huffman code lengths, 37 bits to index an ELQ/KLQ representative, and 40 bits for the AMC quantization indices. These are fixed-length headers and do not include shared parameters or framing overhead.

KLQ has the highest codebook stability for most deletion counts, while the mean KL distortions of KLQ, ELQ, and AMC are close (0.073, 0.076, and 0.081 nats, respectively). AMC has a smaller sampled maximum KL distortion, but its codebook stability is generally lower than KLQ's. All three quantized methods use more bits per sequence than canonical Huffman coding without quantization.

\end{vlong}

\begin{vlong}
\section{Conclusion}
\label{sec:conclusion}

We studied two lattice quantizers on the probability simplex: the Euclidean Lattice Quantizer (ELQ) and the KL Quantizer (KLQ). For both quantizers, we derived KL distortion bounds and quantization stability bounds under perturbations. For KLQ, we also showed that its KL Voronoi cells are determined by lattice-adjacent codewords, which leads to a local-search implementation. We further gave a simple heuristic choice $\alpha^\star\approx 1/(m^2 k\sqrt{2})$ for the lattice offset. Simulations illustrate the trade-off between KL distortion and Huffman code stability, with AMC as a related-work baseline.

\end{vlong}

\begin{vlong}
\appendix
\section{A monotonicity condition for $\alpha$-selection}
\label{app:envelope}

This appendix gives a local monotonicity condition used in the $\alpha$-selection discussion of Section~\ref{sec:elq-alpha-heuristic}.

Fix $k$ and let
$\mathcal W=\{\bw\in\mathbb Z_{\geq 0}^m:\sum_i w_i=km\}$.
For $0<\alpha\leq 1/[m(k+1)]$, set
$\delta=(1/m-\alpha)/k$ and
$\mathcal C_\alpha=\{\alpha\bone+\delta\bw:\bw\in\mathcal W\}$.
For fixed $\bw$, define
$F_\bw(\alpha;\bp)=\sum_i p_i\ln q_i(\alpha;\bw)$, where
$q_i(\alpha;\bw)=\alpha+\delta w_i$, and let
$F^\star(\alpha;\bp)=\max_{\bw\in\mathcal W}F_\bw(\alpha;\bp)$.
Since
$\min_{\bq\in\mathcal C_\alpha}D(\bp\|\bq)
=-H(\bp)-F^\star(\alpha;\bp)$,
a larger $F^\star$ gives a smaller minimum KL distortion.

\begin{lemma}
For fixed $\bw$,
\[
\frac{d}{d\alpha}F_\bw(\alpha;\bp)
=
\frac{1}{1/m-\alpha}
\left(
\frac{1}{m}\sum_i\frac{p_i}{q_i}-1
\right).
\]
\end{lemma}

\begin{proof}
Since
$\frac{d q_i}{d\alpha}=1-w_i/k$,
\[
\frac{d}{d\alpha}F_\bw(\alpha;\bp)
=
\sum_i\frac{p_i}{q_i}\left(1-\frac{w_i}{k}\right).
\]
Using
$1-w_i/k=(1/m-q_i)/(1/m-\alpha)$
and $\sum_i p_i=1$ gives the result.
\end{proof}

\begin{proposition}
\label{prop:envelope-monotone}
Fix $\bp\in\Delta_{m-1}$ and
$0<\alpha_0<1/[m(k+1)]$.
Let
$\bw^\star(\alpha_0)\in
\arg\max_{\bw\in\mathcal W}F_\bw(\alpha_0;\bp)$.
If
\[
\frac{1}{m}
\sum_i
\frac{p_i}
{q_i(\alpha_0;\bw^\star(\alpha_0))}
>1,
\]
then, for all sufficiently small $h>0$,
$F^\star(\alpha_0+h;\bp)>F^\star(\alpha_0;\bp)$.
Hence the minimum KL distortion over $\mathcal C_\alpha$ decreases locally as $\alpha$ increases from $\alpha_0$.
\end{proposition}

\begin{proof}
Let $\bw_0=\bw^\star(\alpha_0)$. The lemma gives
$\frac{d}{d\alpha}F_{\bw_0}(\alpha_0;\bp)>0$,
so for sufficiently small $h>0$,
$F_{\bw_0}(\alpha_0+h;\bp)>F_{\bw_0}(\alpha_0;\bp)$.
Therefore
\[
F^\star(\alpha_0+h;\bp)
\geq
F_{\bw_0}(\alpha_0+h;\bp)
>
F_{\bw_0}(\alpha_0;\bp)
=
F^\star(\alpha_0;\bp).
\]
\end{proof}

By Cauchy--Schwarz,
$\sum_i p_i/q_i\geq 1/\langle\bp,\bq\rangle$.
Hence $\langle\bp,\bq\rangle<1/m$ is a sufficient condition for the derivative in the lemma to be positive. This condition does not hold for every nonuniform $\bp$ and $\bq$, so the proposition gives a local condition rather than global monotonicity in $\alpha$.

On the other hand, the vertex distortion
$B(\alpha)=\ln(1/(1-(m-1)\alpha))$
is strictly increasing in $\alpha$. Thus increasing $\alpha$ can reduce the distortion away from the boundary under the condition above, while increasing the distortion at simplex vertices. This is the trade-off used in Section~\ref{sec:elq-alpha-heuristic}.
\section{KLQ inradius}
\label{app:inradius}

We give the details for Proposition~\ref{prop:klq-inradius}.

\begin{lemma}[Inradius formula]
\label{lem:inradius-formula}
Let $\bn_H(\bq,\bq')$ denote the projection onto
$H_0=\{\bv\in\mathbb{R}^m:\sum_i v_i=0\}$
of the vector $(\ln(q'_i/q_i))_i$. Then
\[
\rho(\bq)
=
\min\left\{
\min_{\bq'\neq\bq}
\frac{D(\bq\|\bq')}
{\|\bn_H(\bq,\bq')\|_2},
\;
\min_i q_i\sqrt{\frac{m}{m-1}}
\right\}.
\]
\end{lemma}

\begin{proof}
We work in the affine hyperplane
$H=\{\bx\in\mathbb{R}^m:\sum_i x_i=1\}$ and write
$\bp=\bq+\bv$ with $\bv\in H_0$.

For $\bq'\neq\bq$, let $a_i=\ln(q'_i/q_i)$. By
Lemma~\ref{lem:kl-convex}, the KL bisector between $\bq$ and $\bq'$
satisfies $\sum_i p_i a_i=0$. Substituting $\bp=\bq+\bv$ gives
$\sum_i v_i a_i=D(\bq\|\bq')$. Since $\bv\in H_0$, the left-hand side
is $\langle\bn_H(\bq,\bq'),\bv\rangle$. Hence the distance from $\bq$
to this bisector is
$D(\bq\|\bq')/\|\bn_H(\bq,\bq')\|_2$.

For the simplex facet $\{p_i=0\}\cap H$, the normal within $H$ is
$\be_i-\bone/m$, whose norm is $\sqrt{(m-1)/m}$. Thus the distance
from $\bq$ to this facet is $q_i\sqrt{m/(m-1)}$. Taking the minimum
over the bisectors and simplex facets gives the result.
\end{proof}

\begin{corollary}[Adjacent-bisector reduction]
\label{cor:klq-adj-binding}
The inradius can be written as
\[
\rho(\bq)
=
\min\left\{
\min_{\bq'\in\mathcal{N}(\bq)}
\frac{D(\bq\|\bq')}
{\|\bn_H(\bq,\bq')\|_2},
\;
\min_i q_i\sqrt{\frac{m}{m-1}}
\right\}.
\]
\end{corollary}

\begin{proof}
By Lemma~\ref{lem:klq-cell-local}, the KL cell is determined by the
lattice-adjacent bisector half-spaces together with the simplex facets.
The inradius is therefore the minimum distance from $\bq$ to these
hyperplanes.
\end{proof}

\begin{lemma}[Inradius lower bound at $\alpha=\delta=1/K$]
\label{lem:rho-lower-bound}
Let $K=m(k+1)$ with $m\ge2$ and $k\ge1$. At
$\alpha=\delta=1/K$,
\[
\rho(m,K)
\triangleq
\min_{\bq\in\mathcal{C}_\delta}\rho(\bq)
\ge
\frac{1-\ln 2}{\ln 2}
\sqrt{\frac{m}{m-1}}\frac{1}{K}.
\]
For fixed $m$, the bound is asymptotically tight as $k\to\infty$.
In particular, $\rho(m,K)=\Theta(1/K)$.
\end{lemma}

\begin{proof}
At $\alpha=\delta=1/K$, write $q_i=r_i/K$, where $r_i\ge1$ and
$\sum_i r_i=K$. By Corollary~\ref{cor:klq-adj-binding}, it is enough
to consider lattice-adjacent codewords
$\bq'=\bq+(\be_a-\be_b)/K$. For this move, $r_b\ge2$.

Let $A=\ln(1+1/r_a)$ and $B=\ln(1+1/(r_b-1))$. Then
$D(\bq\|\bq')=(r_bB-r_aA)/K$. The log-ratio vector has entries
$A$ and $-B$ in coordinates $a$ and $b$, so
$\|\bn_H(\bq,\bq')\|_2^2
=A^2+B^2-(A-B)^2/m$. Therefore the distance to this bisector is
\[
d(r_a,r_b)
=
\frac{1}{K}
\frac{r_bB-r_aA}
{\sqrt{A^2+B^2-(A-B)^2/m}}.
\]

For fixed $r_a$, letting $r_b\to\infty$ gives $B\to0$ and
$r_bB\to1$. Hence
\[
\lim_{r_b\to\infty}d(r_a,r_b)
=
\frac{1}{K}
h(r_a)\sqrt{\frac{m}{m-1}},
\qquad
h(r)
=
\frac{1-r\ln(1+1/r)}
{\ln(1+1/r)}.
\]

The function $h$ is increasing. Indeed, if
$A_r=\ln(1+1/r)$, then $h(r)=1/A_r-r$ and
$h'(r)=1/[r(r+1)A_r^2]-1$. Also,
$\ln(1+1/r)<1/\sqrt{r(r+1)}$ for $r>0$. To see this, set
$z=\sqrt{1+1/r}>1$; the inequality becomes
$2\ln z<z-z^{-1}$, and $z-z^{-1}-2\ln z$ is zero at $z=1$ and
strictly increasing for $z>1$. Thus $h'(r)>0$. Since
$h(1)=(1-\ln2)/\ln2$ and $h(r)\to1/2$, we have
$0<h(r)<1/2$ for finite $r\ge1$.

It remains to compare finite $r_b$ with the limit above. Let
$\beta=h(r_a)$ and set $s=B$. Since
$e^s=r_b/(r_b-1)$, we have $r_bB=s/(1-e^{-s})$. For $s>0$,
$s/(1-e^{-s})\ge1+s/2$. This is equivalent to
$(s-2)e^s+s+2\ge0$; if $L(s)=(s-2)e^s+s+2$, then
$L(0)=L'(0)=0$ and $L''(s)=se^s\ge0$.

Since $1-r_aA=A\beta$, it follows that
$r_bB-r_aA\ge A\beta+s/2$. It is therefore enough to show
\[
\frac{A\beta+s/2}
{\sqrt{A^2+s^2-(A-s)^2/m}}
\ge
\beta\sqrt{\frac{m}{m-1}}.
\]
Squaring and simplifying reduces this to
\[
s\left[
A\beta\left(1-\frac{2\beta}{m-1}\right)
+
s\left(\frac14-\beta^2\right)
\right]
\ge0.
\]
Since $0<\beta<1/2$ and $m\ge2$, both terms in brackets are
nonnegative. Thus
\[
d(r_a,r_b)
\ge
\frac{1}{K}
h(r_a)\sqrt{\frac{m}{m-1}}
\ge
\frac{1}{K}
\frac{1-\ln2}{\ln2}
\sqrt{\frac{m}{m-1}}.
\]

The distance to any simplex facet is at least
$(1/K)\sqrt{m/(m-1)}$. Since $(1-\ln2)/\ln2<1$, the facet term also
satisfies the claimed lower bound.

Finally, consider a codeword with $r_a=1$, $r_b=K-m+1$, and all
other counts equal to $1$. As $k\to\infty$, the distance to the
adjacent bisector obtained by moving one count from $b$ to $a$
approaches
\[
\frac{1}{K}
\frac{1-\ln2}{\ln2}
\sqrt{\frac{m}{m-1}}.
\]
Since the inradius cannot exceed this bisector distance, the lower
bound is asymptotically tight.
\end{proof}
\section{Finite-Type Stability under Deletions}
\label{app:type-stability}

Let $\mathcal T_{L,m}=\{\mathbf c/L:\mathbf c\in\mathbb Z_{\ge0}^m,\ \sum_i c_i=L\}$ be the set of length-$L$ types, and set $d=m-1$. Let $\phi:\Delta_{m-1}\to\mathcal C$ be a quantizer whose closed cells $V(\bq)$ are convex and each contain a Euclidean ball of radius $\rho_\phi>0$ in the affine hull of the simplex. Assume that the cells have disjoint interiors and that all ties are broken by a fixed deterministic rule. Let $X$ have a type $\bp$ that is uniform on $\mathcal T_{L,m}$, and let $Y$ be obtained by deleting $n_{\mathrm{del}}$ symbols, with $1\le n_{\mathrm{del}}<L$. Write $\bp'$ for the type of $Y$. The deleted positions may depend on $X$.

\begin{theorem}[Finite-type stability under deletions]
\label{thm:finite-type-stability}
Using the deletion and codebook stability probabilities defined in Section~\ref{sec:sim}, with a deterministic codebook construction, we have
\begin{equation}
\label{eq:finite-type-stability}
S_{\mathrm{cb}}^\phi(n_{\mathrm{del}})\ge S_{\mathrm{del}}^\phi(n_{\mathrm{del}})\ge\left[1-\frac{2(n_{\mathrm{del}}+\sqrt m)}{L\rho_\phi}\right]_+^{m-1},\qquad [x]_+=\max\{x,0\}.
\end{equation}
\end{theorem}

\begin{proof}
The source model in Section~\ref{sec:sim} satisfies the uniform-type assumption. Indeed, if $\boldsymbol\theta\sim\mathrm{Dir}(1,\ldots,1)$ and $X\mid\boldsymbol\theta$ consists of $L$ i.i.d. symbols, then for any nonnegative count vector $\mathbf c$ with $\sum_i c_i=L$,
\[
\Pr(\bp=\mathbf c/L)=\frac{L!}{\prod_i c_i!}\frac{\Gamma(m)\prod_i\Gamma(c_i+1)}{\Gamma(L+m)}=\binom{L+d}{d}^{-1}.
\]
Thus $\bp$ is uniform over $\mathcal T_{L,m}$. If $\mathbf r$ is the type of the deleted symbols, then $\bp=(1-n_{\mathrm{del}}/L)\bp'+(n_{\mathrm{del}}/L)\mathbf r$, so $\mathrm{TV}(\bp,\bp')\le n_{\mathrm{del}}/L$. Since the positive and negative parts of $\bp-\bp'$ each have total mass $\mathrm{TV}(\bp,\bp')$, we also have
\begin{equation}
\label{eq:finite-type-displacement}
\|\bp-\bp'\|_2\le\sqrt2\,\mathrm{TV}(\bp,\bp')\le\frac{\sqrt2\,n_{\mathrm{del}}}{L}.
\end{equation}

For $u\in\Delta_{m-1}$, define its cell margin by $M_\phi(u)=\mathrm{dist}_2(u,\partial V(\phi(u)))$, with distances taken within the affine hull of $\Delta_{m-1}$. Let $U$ be uniform over the continuous simplex. For $0\le s<\rho_\phi$, contract each cell toward the center of a radius-$\rho_\phi$ ball contained in it by the factor $1-s/\rho_\phi$. Convexity shows that every point in the contracted cell has distance at least $s$ from its boundary, and the volume is multiplied by $(1-s/\rho_\phi)^d$. Applying the argument with a slightly larger margin and taking a limit gives
\begin{equation}
\label{eq:continuous-margin-for-types}
\Pr\bigl[M_\phi(U)>s\bigr]\ge\left[1-\frac{s}{\rho_\phi}\right]_+^d\qquad (s\ge0).
\end{equation}

We next relate continuous volume to the number of types. Consider the full type lattice $\Lambda_L=\{\mathbf z/L:\mathbf z\in\mathbb Z^m,\ \sum_i z_i=L\}$, and let $Q_L(U)$ be a Euclidean nearest point of this lattice, with fixed ties. The nearest point is a type: if its integer coordinate $z_j<0$, then $a_j=z_j-LU_j\le-1$ and, since $\sum_i a_i=0$, some $a_i>0$. Moving one unit from coordinate $i$ to $j$ would change the squared distance by $2(a_j-a_i+1)<0$, a contradiction.

Set $y_i=LU_i$, $f_i=y_i-\lfloor y_i\rfloor$, and $s_0=\sum_i f_i\in\{0,\ldots,m-1\}$. The vertices of $\{v\in[0,1]^m:\sum_i v_i=s_0\}$ are binary vectors with exactly $s_0$ ones. Hence there is a random such vector $B$ with $\mathbb E B_i=f_i$, and
\[
\mathbb E\|B-\mathbf f\|_2^2=\sum_i f_i(1-f_i)\le\frac{s_0(m-s_0)}m\le\frac{\lfloor m^2/4\rfloor}{m}.
\]
Some choice of $B$ therefore rounds $LU$ to a nonnegative integer vector of sum $L$ within Euclidean distance $R_m=\sqrt{\lfloor m^2/4\rfloor/m}$. Since $Q_L(U)$ is a nearest lattice point,
\begin{equation}
\label{eq:type-rounding-radius}
\|U-Q_L(U)\|_2\le\frac{R_m}{L}.
\end{equation}

The affine lattice $\Lambda_L$ has basis $(\be_i-\be_m)/L$, $i=1,\ldots,d$, and fundamental volume $\sqrt m/L^d$. Since the simplex has volume $\sqrt m/d!$, the Voronoi region of any one lattice point contains at most a fraction $d!/L^d$ of the uniform simplex measure. Thus, for any type $v$, $\Pr[Q_L(U)=v]\le d!/L^d$.

Call a length-$L$ type $v$ safe if $M_\phi(v)>\sqrt2\,n_{\mathrm{del}}/L$. By~\eqref{eq:finite-type-displacement}, every safe type keeps the same representative after any deletion of $n_{\mathrm{del}}$ symbols. If $M_\phi(U)>(\sqrt2\,n_{\mathrm{del}}+R_m)/L$, then~\eqref{eq:type-rounding-radius} shows that $Q_L(U)$ is in the same cell as $U$ and is safe. If $N_{\mathrm{safe}}$ is the number of safe types, \eqref{eq:continuous-margin-for-types} and the bound on the mass of one rounded type yield
\[
\frac{N_{\mathrm{safe}}}{\binom{L+d}{d}}\ge a_{L,m}\left[1-\frac{\sqrt2\,n_{\mathrm{del}}+R_m}{L\rho_\phi}\right]_+^d,\qquad a_{L,m}=\frac{L^d}{d!\binom{L+d}{d}}=\prod_{j=1}^{d}\frac{L}{L+j}.
\]
Since $\bp$ is uniform over types, the left-hand side is a lower bound on $S_{\mathrm{del}}^\phi(n_{\mathrm{del}})$.

Finally, $a_{L,m}\ge(L/(L+d))^d$ and $R_m\le\sqrt m/2$. As every cell ball lies in the simplex, $\rho_\phi\le1/\sqrt{m(m-1)}$, so $d\rho_\phi\le1$. Therefore,
\[
S_{\mathrm{del}}^\phi(n_{\mathrm{del}})\ge\left[1-\frac{\sqrt2\,n_{\mathrm{del}}+R_m+d\rho_\phi}{(L+d)\rho_\phi}\right]_+^d\ge\left[1-\frac{2(n_{\mathrm{del}}+\sqrt m)}{L\rho_\phi}\right]_+^d.
\]
Equal representatives give equal codebooks under deterministic coding, so $S_{\mathrm{cb}}^\phi(n_{\mathrm{del}})\ge S_{\mathrm{del}}^\phi(n_{\mathrm{del}})$, completing the proof.
\end{proof}

For ELQ and KLQ, the convex-cell assumption and the required radius follow from their cell descriptions and Propositions~\ref{prop:elq-inradius} and~\ref{prop:klq-inradius}.

\end{vlong}

\Section{References}
\bibliographystyle{IEEEtran}
\bibliography{reference}

@INPROCEEDINGS{ITWpaper,
  author={Luo, Haoxuan and Farnoud, Farzad},
  booktitle={2025 IEEE Information Theory Workshop (ITW)}, 
  title={Improving Smoothness in {Huffman} Coding: Canonical and Pre-allocated Variations}, 
  year={2025},
  volume={},
  number={},
  pages={602--607},
  doi={10.1109/ITW62417.2025.11240326}}

@inproceedings{montanari2008smooth,
  title={Smooth compression, {Gallager} bound and nonlinear sparse-graph codes},
  author={Montanari, Andrea and Mossel, Elchanan},
  booktitle={2008 IEEE International Symposium on Information Theory},
  pages={2474--2478},
  year={2008}
}

@article{Huffman,
    author   = {Huffman, David A.},
    journal  = {Proceedings of the IRE},
    title    = {A Method for the Construction of Minimum-Redundancy Codes},
    year     = {1952},
    volume   = {40},
    number   = {9},
    pages    = {1098-1101},
    doi      = {10.1109/JRPROC.1952.273898}
}

@article{arithmetic,
    title     = {Arithmetic coding for data compression},
    author    = {Witten, Ian H and Neal, Radford M and Cleary, John G},
    journal   = {Communications of the ACM},
    volume    = {30},
    number    = {6},
    pages     = {520--540},
    year      = {1987},
    publisher = {ACM New York, NY, USA}
}

@incollection{synchronization,
  title={Algorithms for delta compression and remote file synchronization},
  author={Suel, Torsten and Memon, Nasir},
  booktitle={Lossless Compression Handbook},
  editor={Sayood, Khalid},
  pages={269--290},
  year={2002},
  publisher={Academic Press}
}

@techreport{rsync,
  title={The rsync algorithm},
  author={Tridgell, Andrew and Mackerras, Paul},
  institution={Australian National University},
  number={TR-CS-96-05},
  year={1996}
}

@book{conway1999sphere,
  author    = {Conway, John H. and Sloane, Neil J. A.},
  title     = {Sphere Packings, Lattices and Groups},
  edition   = {3rd},
  address   = {New York},
  publisher = {Springer},
  year      = {1999}
}

@article{lloyd1982least,
  title={Least squares quantization in {PCM}},
  author={Lloyd, Stuart},
  journal={IEEE Transactions on Information Theory},
  volume={28},
  number={2},
  pages={129--137},
  year={1982},
  publisher={IEEE}
}

@article{JMLR:v6:banerjee05b,
  author  = {Arindam Banerjee and Srujana Merugu and Inderjit S. Dhillon and Joydeep Ghosh},
  title   = {Clustering with {Bregman} Divergences},
  journal = {Journal of Machine Learning Research},
  year    = {2005},
  volume  = {6},
  number  = {58},
  pages   = {1705--1749},
  url     = {http://jmlr.org/papers/v6/banerjee05b.html}
}

@ARTICLE{vatedka2020local,
  author={Vatedka, Shashank and Tchamkerten, Aslan},
  journal={IEEE Transactions on Information Theory}, 
  title={Local Decode and Update for Big Data Compression}, 
  year={2020},
  volume={66},
  number={9},
  pages={5790-5805},
  doi={10.1109/TIT.2020.2999909}}

@INPROCEEDINGS{varshney2009malleable,
  author={Varshney, Lav R. and Kusuma, Julius and Goyal, Vivek K.},
  booktitle={2009 IEEE International Symposium on Information Theory}, 
  title={Malleable coding with edit-distance cost}, 
  year={2009},
  volume={},
  number={},
  pages={204--208},
  doi={10.1109/ISIT.2009.5205494}}

@INPROCEEDINGS{ISITpaper,
  author={Luo, Haoxuan and Li, Yuting and Farnoud, Farzad and Gabrys, Ryan},
  booktitle={2026 IEEE International Symposium on Information Theory (ISIT)}, 
  title={Covering Codes with Low {KL} Distortion for Smooth Compression}, 
  year={2026},
  volume={},
  number={},
  pages={1--6},
  doi={10.1109/ISIT62367.2026.11653662}}

@ARTICLE{adler2023efficient,
  author={Adler, Aviv and Tang, Jennifer and Polyanskiy, Yury},
  journal={IEEE Journal on Selected Areas in Information Theory}, 
  title={Efficient Representation of Large-Alphabet Probability Distributions}, 
  year={2022},
  volume={3},
  number={4},
  pages={651--663},
  doi={10.1109/JSAIT.2023.3234502}}

@article{geiger2016greedy,
  author  = {Geiger, Bernhard C. and B{\"o}cherer, Georg},
  title   = {Greedy Algorithms for Optimal Distribution Approximation},
  journal = {Entropy},
  volume  = {18},
  number  = {7},
  year    = {2016},
  note    = {Art. no. 262},
  doi     = {10.3390/e18070262}
}

@INPROCEEDINGS{reznik2011,
  author={Reznik, Yuriy A.},
  booktitle={2011 Data Compression Conference}, 
  title={An Algorithm for Quantization of Discrete Probability Distributions}, 
  year={2011},
  volume={},
  number={},
  pages={333--342},
  doi={10.1109/DCC.2011.40}}

\end{document}